\documentclass[bibyear]{aa}  
\usepackage{natbib}
\usepackage{graphicx}
\usepackage{dcolumn}
\usepackage{rotating}
\usepackage{booktabs}

\usepackage{txfonts}
\usepackage{subcaption}
\usepackage{caption}
\usepackage{color}
\usepackage{natbib,twoopt}
\usepackage[breaklinks,draft=false]{hyperref}
\bibpunct{(}{)}{;}{a}{}{,}             
\definecolor{cobalt}{rgb}{0.06, 0.2, 0.65}
\hypersetup{
  colorlinks=true,
  citecolor=cobalt,
  linkcolor=cobalt,
  urlcolor=cobalt
}

\usepackage{twoopt}
\makeatletter
  \newcommandtwoopt{\citeads}[3][][]{\href{http://adsabs.harvard.edu/abs/#3}%
    {\def\hyper@linkstart##1##2{}%
     \let\hyper@linkend\@empty\citealp[#1][#2]{#3}}}
  \newcommandtwoopt{\citepads}[3][][]{\href{http://adsabs.harvard.edu/abs/#3}%
    {\def\hyper@linkstart##1##2{}%
     \let\hyper@linkend\@empty\citep[#1][#2]{#3}}}
  \newcommandtwoopt{\citetads}[3][][]{\href{http://adsabs.harvard.edu/abs/#3}%
    {\def\hyper@linkstart##1##2{}%
     \let\hyper@linkend\@empty\citet[#1][#2]{#3}}}
  \newcommandtwoopt{\citeyearads}[3][][]%
    {\href{http://adsabs.harvard.edu/abs/#3}
    {\def\hyper@linkstart##1##2{}%
     \let\hyper@linkend\@empty\citeyear[#1][#2]{#3}}}
\makeatother

\def\ts     {\thinspace} 
\usepackage{amsmath}

\def\kms  {\ifmmode{{\rm \ts km\ts s}^{-1}}\else{\ts km\ts s$^{-1}$\ts}\fi}
\def\msol {\ifmmode{{\rm M}_{\odot}}\else{M$_{\odot}$\ts}\fi}
\def\cii  {\ifmmode{{\rm [C}{\rm \scriptstyle II}]}\else{[C\ts {\scriptsize II}]\ts}\fi}
\def\ci   {\ifmmode{{\rm [C}{\rm \scriptstyle I}]}\else{[C\ts {\scriptsize I}]\ts}\fi}
\def\m    {\ifmmode{\mu {\rm m}}\else{$\mu$m}\fi}
\def\hi   {\ifmmode{{\rm H}{\rm \scriptstyle I}}\else{H\ts {\scriptsize I}\ts}\fi}
\def\hii  {\ifmmode{{\rm H}{\rm \scriptstyle II}}\else{H\ts {\scriptsize II}\ts}\fi}
\def\nii  {\ifmmode{{\rm [N}{\rm \scriptstyle II}]}\else{[N\ts {\scriptsize II}]\ts}\fi}
\def\oiii {\ifmmode{{\rm [O}{\rm \scriptstyle III}]}\else{[O\ts {\scriptsize III}]\ts}\fi}
\def\hh   {\ifmmode{{\rm H}_2}\else{H$_2$\ts}\fi}
\def\nhh  {\ifmmode{N({\rm H}_2)}\else{$N$(H$_2$)\ts}\fi}
\def\microns {\ifmmode{\mu{\rm m}}\else{$\mu$m\ts}\fi}

\def\lya {\ifmmode{{\rm Ly}{\alpha}}\else{Ly$\alpha$\ts}\fi}
\def\ha   {\ifmmode{{\rm H}{\alpha}}\else{H$\alpha$\ts}\fi}
\def\hb   {\ifmmode{{\rm H}{\beta}}\else{H$\beta$\ts}\fi}

\def\ergcms   {\ifmmode{{\rm erg}\ts{\rm cm ^{-2}}\ts{\rm s ^{-1}}}\else{${\rm erg}\ts{\rm cm ^{-2}}\ts{\rm s ^{-1}}$\ts}\fi}
\begin{document}

   \title{The Extended Mapping Obscuration to Reionization with ALMA (Ex-MORA) Survey: A Molecular Gas Line Search}
   \titlerunning{Ex-MORA Survey: A Molecular Gas Line Search}

   \author{V. Cat\'an
          \inst{1,2}
          \and
          J. Gonz\'alez-L\'opez\inst{1,2,3}
          \and
          M. Aravena\inst{4,2}
          \and A.S. Long\inst{5}
          \and C.M. Casey\inst{6,7}
          \and D.L. Clements\inst{8} 
          \and E. da Cunha\inst{9}
          \and M. Franco \inst{10}
          \and A.M. Koekemoer\inst{11}
           \and A. Man \inst{12}
          \and J. Spilker\inst{13}
          \and  E. Treister\inst{14}
          }

   \institute{Instituto de Astrof\'isica, Facultad de F\'isica, Pontificia Universidad Cat\'olica de Chile, Av. Vicuña Mackenna 4860, 782-0436 Macul, Santiago, Chile\\
   \email{victoria.catan@uc.cl}
   \and Millennium Nucleus for Galaxies (MINGAL)
           \and
             Las Campanas Observatory, Carnegie Institution of Washington,
  Ra\'ul Bitr\'an 1200, La Serena, Chile
             \and Instituto de Estudios Astrofísicos, Facultad de Ingenier\'ia y Ciencias, Universidad Diego Portales, Av. Ej\'ercito Libertador 441, Santiago, Chile
            \and Department of Astronomy, University of Washington, Seattle, WA 98195, USA
             \and Department of Physics, University of California, Santa Barbara, Santa Barbara, CA 93106, USA 
             \and Cosmic Dawn Center (DAWN), Denmark 
             \and Blackett Lab, Imperial College London, Prince Consort Road, London SW7 2AZ, UK
             \and International Centre for Radio Astronomy Research (ICRAR), University of Western Australia, Crawley, WA 6009, Australia
             \and  Université Paris-Saclay, Université Paris Cité, CEA, CNRS, AIM, 91191, Gif-sur-Yvette, France
             \and Space Telescope Science Institute, 3700 San Martin Drive, Baltimore, MD 21218, USA
             \and Department of Physics \& Astronomy, University of British Columbia, 6224 Agricultural Road, Vancouver, BC V6T 1Z1, Canada
\and Department of Physics and Astronomy, Texas A\&M University, 4242 TAMU, College Station, TX 77843-4242, USA
\and Instituto de Alta Investigación, Universidad de Tarapacá, Casilla 7D, Arica, Chile}

 
\abstract
{Current investigations of cold gas in the interstellar medium are limited by selection biases, as most samples are derived from optical or near-infrared surveys that preferentially select massive, low-obscuration galaxies. While previous research has established key scaling relations for the main sequence (MS), these studies often exclude gas-rich passive systems and inefficient star-forming populations that are not identified by conventional color-selection criteria. Overcoming these limitations requires a large-volume molecular-line survey that identifies galaxies based on their molecular gas reservoirs rather than stellar luminosity.}
{This study provides a census of molecular gas content in galaxies spanning a broad cosmic range ($0.5 < z < 5$) through a spectral search within the Extended Mapping of Obscuration to Reionization Survey (Ex-MORA) ALMA Band 4 observations.}
{The spectral axis of ALMA data in the Ex-MORA field, covering 577 arcmin$^2$, enabled a dual-strategy molecular line search: an unbiased search for high-significance line emitters and a targeted search utilizing spectroscopic redshift priors. In total, 52 galaxies were detected in CO or \ci line transitions, with two exhibiting two distinct lines. Physical properties were derived using \texttt{CIGALE} with non-parametric star formation history modeling, and molecular gas masses were estimated using metallicity-dependent conversion factors.}
{Our sample is dominated by MS systems, characterized by high stellar masses (median $\log M_*/\text{M}_{\odot} = 10.83$) and substantial molecular gas reservoirs ($>10^{10}\,\text{M}_{\odot}$). The gas fraction ($\mu_{\text{gas}}$) increases with redshift among active populations; however, this trend may be influenced by environmental factors, particularly a significant bias at high redshift due to a gas-rich protocluster at $z \approx 2.4$. In contrast, the overdensity at $z \approx 0.73$ is indistinguishable from the field population and does not bias the results. Five transitional Green Valley sources are identified, exhibiting a peak in $\mu_{\text{gas}}$ at cosmic noon, potentially due to stellar feedback during this epoch that suppresses star formation by heating the gas rather than immediate reservoir depletion. Trying to conclude that this process happens to all Green Valley galaxies is beyond the scope and capabilities of our survey, but our results show that at least some GV galaxies could be affected by feedback.  Kinematic analysis of resolved sources indicates a predominance of rotating disk structures. }
{Large-area line searches are highly effective at recovering gas-massive galaxies across diverse evolutionary states, including transitional phases often missed by traditional surveys. The prevalence of MS systems and high gas masses underscores this approach's ability to identify and characterize the most prominent gas reservoirs.}

\maketitle
\section{Introduction}
A critical aspect of the current understanding of galaxy evolution has been the determination of the so-called star-forming main sequence (MS). Observed from the local universe up to $z \approx 6$, this tight correlation between stellar mass ($M_*$) and star formation rate (SFR) defines the MS of typical star-forming galaxies, usually distinguishing them from the starbursts (SB) positioned above it and the passive or quiescent systems below \citep[e.g.][]{popessoMainSequenceStarforming2023, speagleHighlyConsistentFramework2014,noeskeStarFormationAEGIS2007}. This evolutionary progression is fundamentally regulated by the availability and physical state of molecular gas reservoirs ($M_{\text{mol}}$). Recent surveys have shown that the molecular gas-to-stellar mass ratio, or gas fraction ($\mu_{\text{gas}} = M_{mol}/M_*$), increases significantly with redshift, suggesting that the higher SFR density in the early universe was mainly driven by the greater abundance of cold gas and not by a fundamental change in star-formation efficiency \citep{tacconiPHIBSSUnifiedScaling2018,walterEvolutionBaryonsAssociated2020a}.

Despite its importance, measuring the gas reservoirs of galaxies remains challenging, as molecular hydrogen ($H_2$), the primary constituent of the cold interstellar medium (ISM), lacks a permanent dipole moment. As its lowest allowed quadrupole rotational transitions require excitation energies of ~500 K above the ground state, it remains unexcited in the 10–50 K cold ISM, rendering the molecule effectively undetectable. Consequently, most studies rely on indirect tracers, each with specific physical assumptions and uncertainties \citep{saintongeColdInterstellarMedium2022,hodgeHighredshiftStarFormation2020a,carilliCoolGasHigh2013,bolattoCOtoH2ConversionFactor2013a}.

The most widely used proxy is carbon monoxide (CO), the second-most abundant molecule, with rotational transitions that are relatively easy to detect. However, converting CO luminosity into molecular gas mass requires the $\alpha_{\text{CO}}$ conversion factor, which varies significantly with gas-phase metallicity and radiation field intensity \citep{saintongeColdInterstellarMedium2022,hodgeHighredshiftStarFormation2020a,carilliCoolGasHigh2013,bolattoCOtoH2ConversionFactor2013a}. Additionally, in low-metallicity or high-radiation environments, CO can be photodissociated into atomic carbon (CI) while the $H_2$ core remains self-shielded \citep[e.g.][]{bisbasEffectiveDestructionCO2015,bisbasCosmicrayInducedDestruction2017}. This leads to the presence of the denominated "CO-dark" gas, which is missed by traditional CO surveys. Due to its fine-structure lines, often optically thin and capable of tracing these otherwise invisible reservoirs across a wide range of physical conditions, \ci emission has emerged as a robust yet less bright alternative. \citep{saintongeColdInterstellarMedium2022,hodgeHighredshiftStarFormation2020a,carilliCoolGasHigh2013,papadopoulosCILinesTracers2004a}.

Most current studies of molecular gas are subject to selection bias, as galaxies are often preselected from optical or near-infrared surveys that favor massive or low-obscuration systems. As a result, programs like PHIBSS \citep{tacconiPHIBSSUnifiedScaling2018} have established essential scaling relations for MS galaxies; yet they often miss fainter populations or those with unusual gas-to-SFR ratios \citep{solimanoMolecularGasBudget2021b,catanMolecularGasBudget2024}. To address these limitations, unbiased line searches have emerged as a robust alternative (e.g., ALMA Spectroscopic Survey in the Hubble Ultra Deep Field (ASPECS) \citep{walterALMASpectroscopicSurvey2016, aravenaALMASpectroscopicSurvey2019}, The Atacama Large Millimeter/submillimeter Array (ALMA) Calibrators (ALMACAL) \citep{oteoALMACALFirstDualband2016,hamanowiczALMACALVIIIPilot2023a}). By scanning entire frequency bands without prior target selection, these surveys can identify gas-rich passive galaxies and inefficient star-forming systems that escape traditional color- or continuum-selected surveys. Nevertheless, their reach is often limited by small spatial footprints.

An alternative approach was recently explored in \cite{naritaALMALensingCluster2026}, which applies unbiased searches to previously studied ALMA continuum surveys. While continuum studies usually collapse these cubes into 2D images, the full spectral axis remains a rich resource for identifying emission lines (CO, \ci, and \cii) across specific redshift windows. Using this method, \cite{naritaALMALensingCluster2026} found seven confirmed line emitters. Similarly to this work, we employ the same methodology: repurposing large-area ALMA data cubes from the Mapping Obscuration to Reionization with ALMA survey (MORA; \citep{caseyMappingObscurationReionization2021a,zavalaEvolutionIRLuminosity2021}) and its expansion (Ex-MORA; \citep{longExtendedMappingObscuration2024}) as contiguous, high-volume line searches. By searching a total of 577 arcmin$^2$ within the COSMOS field \citep{scovilleCosmicEvolutionSurvey2007}, we probe a much larger cosmic volume than traditional pencil-beam surveys. This approach enables us to construct a line-selected sample and investigate the properties of the cold gas in some of the most massive and extreme systems in the early universe. 

In this context, we identified and studied the molecular gas properties of 52 line-selected galaxies and compared them with established scaling relations. In Section \ref{sec:2}, we present the observations and auxiliary data used. Section \ref{sec:3} describes the data reduction. Section \ref{sec:4} presents our line selection technique, separating unbiased from targeted detections. In Section \ref{sect:SED}, we present our spectral energy distribution (SED) fitting and the properties derived. Section \ref{sec:6} summarizes the method used for our molecular gas estimates. Section \ref{sec:7} presents our results and their comparisons with previous studies, which are then discussed in Section \ref{sec:8}. Finally, we summarize and conclude our work in Section \ref{sec:9}. Throughout this analysis, we adopt a standard flat $\Lambda$CDM cosmological framework. We use a Hubble constant of $H_0 = 70$ km s$^{-1}$ Mpc$^{-1}$, a dark energy density parameter of $\Omega_{\Lambda} = 0.7$, and a matter density of $\Omega_{M} = 0.3$. Physical galaxy properties, including stellar masses and star formation rates, are derived assuming a \cite{chabrierGalacticStellarSubstellar2003a} Initial Mass Function (IMF).

\section{Data and observations} \label{sec:2}

\subsection{ALMA observations: The MORA and Ex-MORA surveys}

We use observational data from the MORA survey (ALMA Project ID: 2018.1.00231.S; PI: C.M. Casey; \citep{caseyMappingObscurationReionization2021a,zavalaEvolutionIRLuminosity2021}) and its expansion, the Ex-MORA survey (ALMA Project ID: 2021.1.00225.S; PI: C.M. Casey; \citep{longExtendedMappingObscuration2024}). Both blank-field surveys employ ALMA Band 4 (2 mm) to identify high-redshift ($z > 3$) dusty star-forming galaxies (DSFGs). In this study, we used the spectral dimension of these observations to search for line emission.

The MORA observations used ALMA’s C43-3 configuration, providing a beam size of $1".83 \times 1".43$, sufficient to encompass entire galaxies and prevent flux dilution. Observations were conducted from March 27 to April 3, 2019, under average atmospheric conditions with a precipitable water vapor (PWV) of 5 mm, covering 184 arcmin$^2$. The frequency setup included two tunings: Tuning 139 (MORA 1), centered at 139.03 GHz, and Tuning 147 (MORA 2), centered at 147.28 GHz, each with a 7.5 GHz bandwidth.

\begin{figure}
    \centering
    \includegraphics[width=\linewidth]{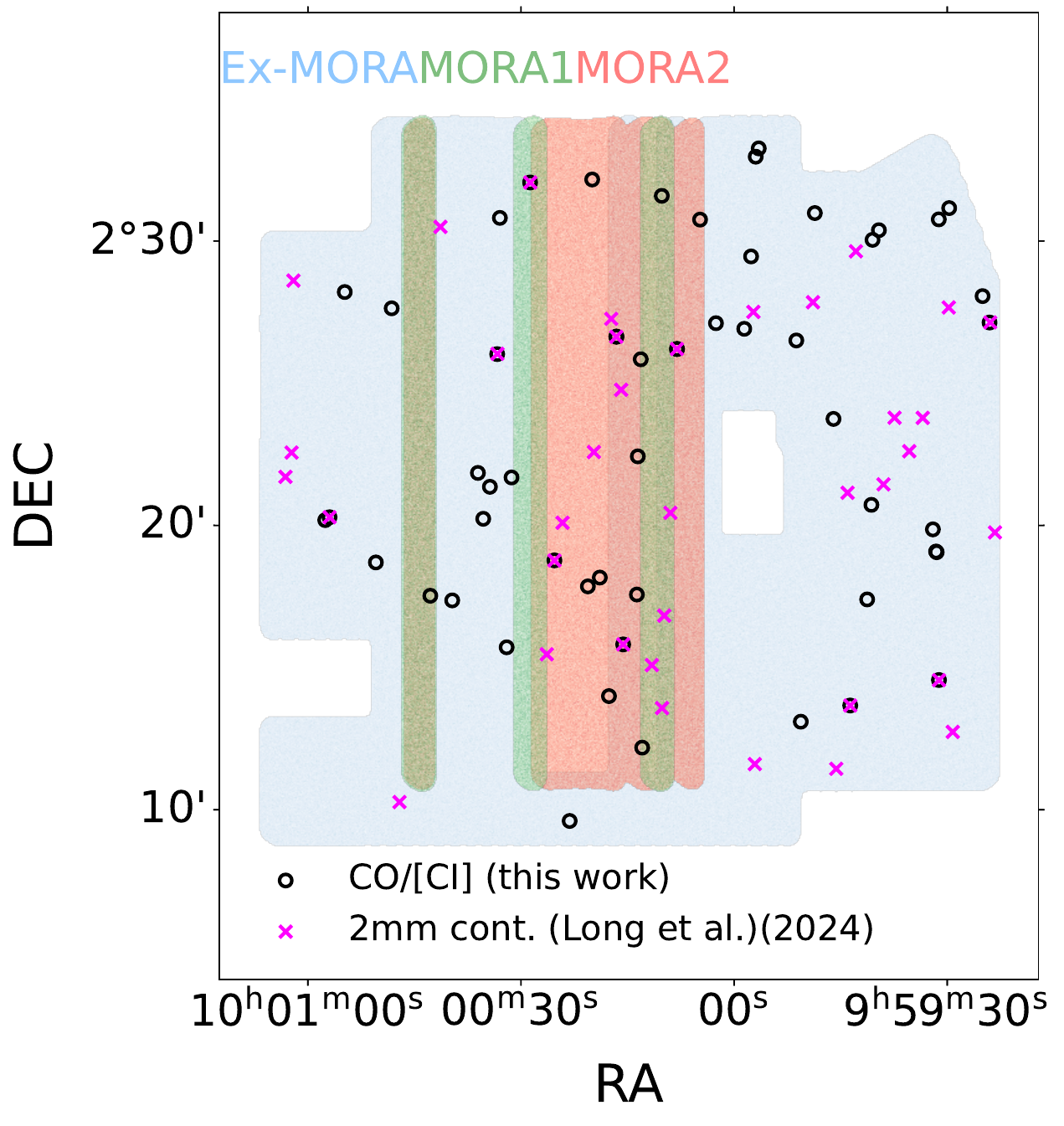}
    \caption{The colored regions indicate the coverage of the Ex-MORA (blue), MORA1 (green), and MORA2 (orange) survey footprints. Black open circles denote the positions of CO and/or \ci line detections identified in this work. For comparison, magenta crosses indicate the 2mm continuum sources reported by \cite{longExtendedMappingObscuration2024}.}
    \label{fig:detections}
\end{figure}

The Ex-MORA survey expanded the original MORA footprint from approximately $0.05\text{ deg}^2$ to $0.2\text{ deg}^2$, though only partial completion was achieved, resulting in a final mosaic covering $577\text{ arcmin}^2$. This expansion increased spatial coverage by nearly a factor of three, enhancing the survey's ability to sample diverse galaxy populations. Data were observed from April 2022 to January 2023 using ALMA's C43-2 configuration. The technical specifications, including the observing frequency (139.03 GHz) and bandwidth, matched those of MORA 1 to enable seamless data merging. Observations were conducted under low PWV ($1.10\text{ mm}$) conditions. Table~\ref{table: survey} summarizes the survey parameters, and Figure~\ref{fig:detections} shows the combined survey coverage and detections for this work, as well as the continuum detections from \cite{longExtendedMappingObscuration2024}.

\begin{table*}

\centering 
\caption{Summary of survey configurations and sensitivities. }
\begin{tabular}{lcccccc c}
\hline\hline
Survey & Configuration & Coverage   & Beam           & Mean           & Spectral  & Pointings &Central \\
 &  &    & size          & RMS           &  resolution & &Frequencies \\
       &               & [arcmin$^2$] &                        & [$\mu$ Jy beam$^{-1}$] & [GHz]                &       &  [GHz]  \\
\hline
MORA   & C43-3         & 184        & $1".83 \times1".43$  & 82.5               & 0.016               & 2086     & 139.03 , 147.28\\
EX-MORA & C43-2         & 577        & $1".68 \times 1".44$ & 89.12             & 0.016               & 4851    &  139.03\\
\hline
\hline
\end{tabular}
\tablefoot{The table lists the ALMA configurations, total sky coverage, synthesized beam sizes, mean RMS sensitivity for each survey pointing, spectral resolution, the total number of pointings, and the central frequencies for the MORA and EXMORA programs.}
\label{table: survey}
\end{table*}

\begin{figure}
    \centering
    \includegraphics[width=\linewidth]{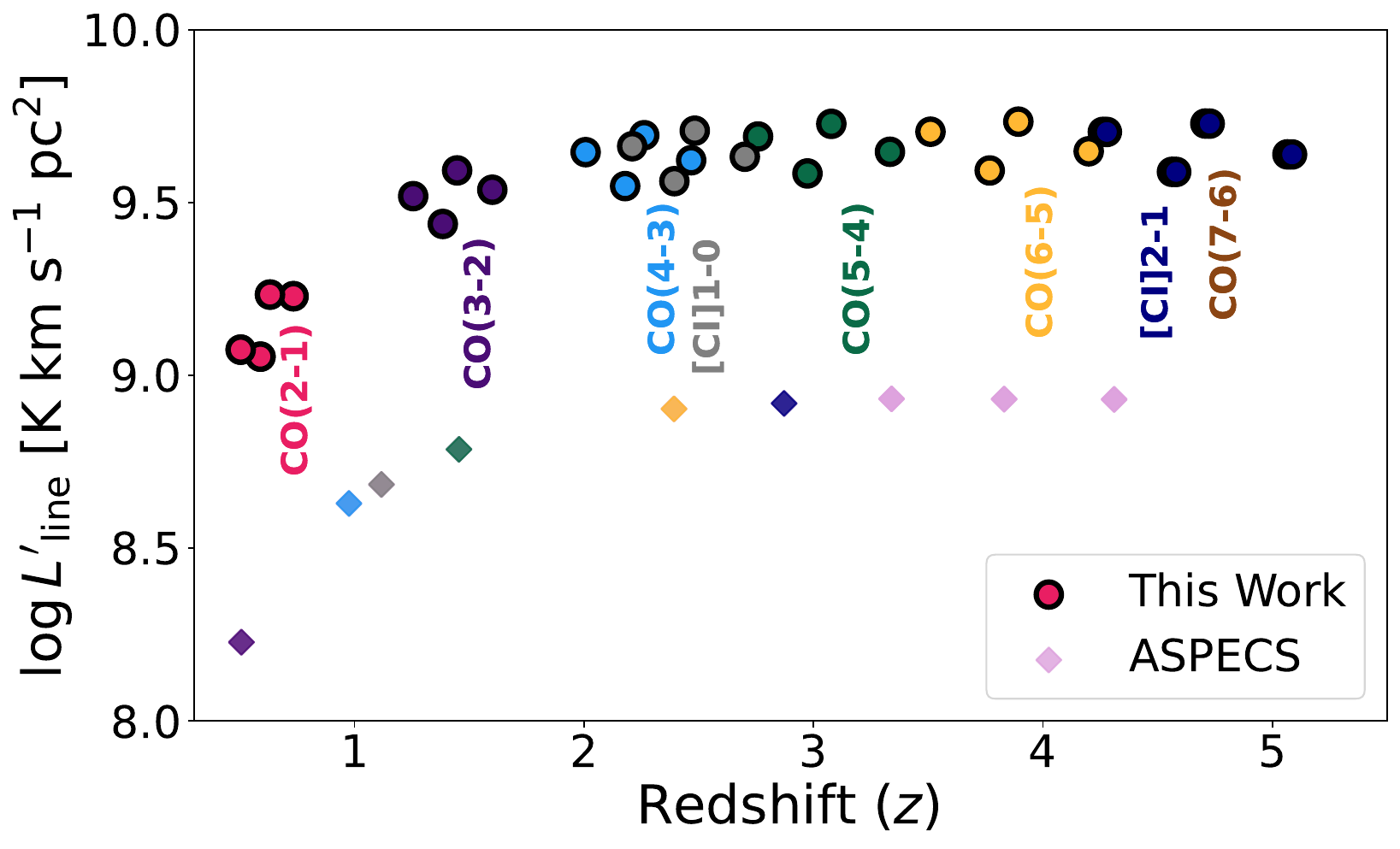}
    \caption{Line luminosity limits for our sample as a function of redshift. We assume a 5$\sigma$ limit for a line width of 200 kms$^{-1}$. The circles highlight the median sensitivity limit across the bandwidth. The diamonds correspond to the limits retrieved from the ASPECS sample \citep{decarliALMASpectroscopicSurvey2020}}
    \label{fig:luminosities}
\end{figure}

Figure \ref{fig:luminosities} presents the line-luminosity limits for our sample and compares them with those of ASPECS \citep{decarliALMASpectroscopicSurvey2020}. These limits represent the median sensitivity for a $5\sigma$ line search. Our unbiased detection limits are approximately 1 dex higher than those of the ASPECS deep fields, leading to a bias toward more massive, gas-rich systems. However, by complementing with known spectroscopic redshifts, our dual search may recover fainter emissions, potentially revealing galaxies with lower molecular gas masses.

 \subsection{Auxiliary data}

\subsubsection{COSMOS-Web catalog}

A key advantage of Ex-MORA is its location within the COSMOS field, a 2 deg$^2$ region near the celestial equator \citep{scovilleCosmicEvolutionSurvey2007}. This area provides extensive multi-wavelength observations from X-rays to radio frequencies. The COSMOS-Web survey recently contributed 255 hours of observations in four JWST NIRCam (F115W, F150W, F277W, F444W) and one JWST MIRI (F770W) bands \citep{caseyCOSMOSWebOverviewJWST2023, shuntovCOSMOS2025COSMOSWebGalaxy2025,harishCOSMOSWebMIRIData2025,francoPhysicalPropertiesGalaxies2025}. The survey, centered within COSMOS, covers $0.54 \text{ deg}^2$ in NIRCam and $0.19 \text{ deg}^2$ in MIRI. The final mosaic measures $46' \times 46'$ with a pixel scale of $0.03''$ per pixel \citep{shuntovCOSMOS2025COSMOSWebGalaxy2025}. Based on these data, \cite{shuntovCOSMOS2025COSMOSWebGalaxy2025} released the first COSMOS-Web (DR1) catalog, which includes measurements and properties for over 700,000 galaxies.

For our multi-wavelength analysis and sample characterization, we used the COSMOS-Web DR1 catalog described by \cite{shuntovCOSMOS2025COSMOSWebGalaxy2025}, employing photometric redshifts ($z_{\text{phot}}$) to initially identify candidates. We also relied on the catalog’s model-based photometry, particularly for spectral energy distribution (SED) fitting. The filters used for SED fitting included JWST/NIRCam F115W, F150W, F277W, F444W, JWST/MIRI F770W, HST/ACS F814W \citep{koekemoerCOSMOSSurveyHubble2007}, Subaru/Suprime-Cam $g+$, $r+$, $i+$, $z+$ \citep{taniguchiSubaruCOSMOS202015}, and VISTA/VIRCAM $J$, $H$, $K_s$ \citep{mccrackenUltraVISTANewUltradeep2012}.

\subsubsection{COSMOS spectroscopic catalog}

To complement the photometric catalogs, we used the spectroscopic redshift compilation by \citet{khostovanCOSMOSSpectroscopicRedshift2025}, which spans redshifts from the local universe up to 8 and is classified using quality flags. This compilation combined spectroscopic redshift information from 138 individual programs covering a $10\text{ deg}^2$ area, providing redshifts for 266,284 unique sources. The redshifts were obtained by systematically mining archival databases, including the NASA/IPAC Infrared Science Archive (IRSA), the ESO Science Portal, and CDS/VizieR, as well as survey websites, literature reviews, and direct communication with principal investigators.

In this study, we used the catalog to identify counterparts to our sources and classify emission lines using a 0.6" crossmatch radius. To ensure a high-fidelity sample for validating our line search, we restricted our analysis to galaxies with high-confidence quality flags: only redshifts with a confidence level exceeding 85\%.

\section{Data reduction} \label{sec:3}

We processed the raw interferometric visibilities using the Common Astronomy Software Applications (CASA) software package version 6.5.4-9 and the official ALMA/VLA pipeline version 2023.1.0.125 \citep[CASA;][]{teamCASACommonAstronomy2022}. Both surveys consist of large mosaics with multiple pointings in ALMA Band 4, which we co-added into contiguous spectral cubes. The cubes were optimized for the detection of faint, previously unknown emission lines through line searches, employing the \texttt{tclean} task for imaging and deconvolution. To ensure consistency across the combined mosaic, we fully automated and parameterized the process. Surface brightness sensitivity was prioritized over angular resolution by applying a natural weighting scheme (\texttt{weighting='natural'}) to the visibility data. Favoring shorter baselines maximizes point-source sensitivity and enhances the signal-to-noise ratio (SNR) for extended emission.

To reconstruct the large area, we set the \texttt{gridder} parameter to \texttt{mosaic}. This mode applies the primary beam response for each pointing to weight overlapping regions, ensuring a uniform noise floor and accurate flux calibration across the mosaic. For deconvolution, we employed the auto-multithresh algorithm, which is advantageous for unbiased searches as it automatically masks regions based on SNR thresholds, sidelobe levels, and noise characteristics. Restricting deconvolution to regions with high-fidelity astrophysical emission effectively prevented noise artifacts in the final data products \citep{ teamCASACommonAstronomy2022}.

The ALMA receivers observe both the Lower Sideband (LSB) and Upper Sideband (USB) simultaneously; therefore, we generated independent spectral cubes for each sideband. The MORA 1 (Tuning 139) and MORA 2 (Tuning 147) tunings were processed as separate measurement sets. The same pipeline was applied to the Ex-MORA data to ensure a homogeneous dataset. To achieve maximum velocity resolution for line fitting, we produced the final data cubes at the native spectral resolution of 16 MHz.

\section{Catalog Selection} \label{sec:4}
Two independent, complementary methods were used to construct the line-emitter galaxy sample, aiming to maximize both reliability and depth in the final catalog. The first approach involved an unbiased search of the ALMA spectral cubes, which identified high-significance emission lines without prior knowledge of source positions or redshifts and uncovered galaxies with high equivalent widths. To complement this, a targeted search was conducted using spectroscopic redshifts, enabling the detection of fainter emission at expected frequencies where an unbiased algorithm might not reach sufficient statistical significance. By focusing on galaxies with pre-determined redshifts, the SNR threshold for detection was lowered while maintaining high confidence in line identification. This dual strategy enables the probing of both luminous starbursts and fainter, gas-rich systems, providing a more comprehensive view of the ISM across cosmic time.

\subsection{Unbiased Search}

Emission lines were identified without the bias of prior spatial or spectral information using the \texttt{Lineseeker} algorithm  detailed in \citep{gonzalez-lopezAtacamaLargeMillimeter2019}, which is specifically designed to process large ALMA data cubes. We present a brief explanation of the method in Appendix \ref{app: noise}, for more detail refer to \citep{gonzalez-lopezAtacamaLargeMillimeter2019}. The algorithm identified candidates and assigned each a \texttt{PPessimistic} value, a conservative fidelity parameter. The raw p-value represents the probability that a detection is due to random noise, derived by comparing positive detections against negative noise spikes in the same cube. The \texttt{PPessimistic} value applies a correction under the assumption that each spectral convolution is independent. Since these convolutions are not strictly independent, the \texttt{PPessimistic} value serves as a lower limit on fidelity, with the true significance lying between the original p-value and this pessimistic estimate. A \texttt{PPessimistic} value of 0 indicates a highly reliable detection with no corresponding negative features at that significance, while values approaching 1 are statistically indistinguishable from noise. To ensure a high-purity catalog, an SNR threshold of 6.0 or higher was applied, and the sample was restricted to candidates with \texttt{PPessimistic} $< 0.5$. These criteria ensured a reliable catalog and reduced false positives to under 6\%, as indicated by the p-values from \texttt{Lineseeker}. The initial selection was cross-matched within a 0.6" radius to the COSMOS-Web DR1 and the auxiliary spectroscopic catalog. No bright lines lacking a multi-wavelength counterpart were found; every unbiased candidate possessed one, providing independent validation of the line.

\begin{figure*}
\includegraphics[width=\textwidth]{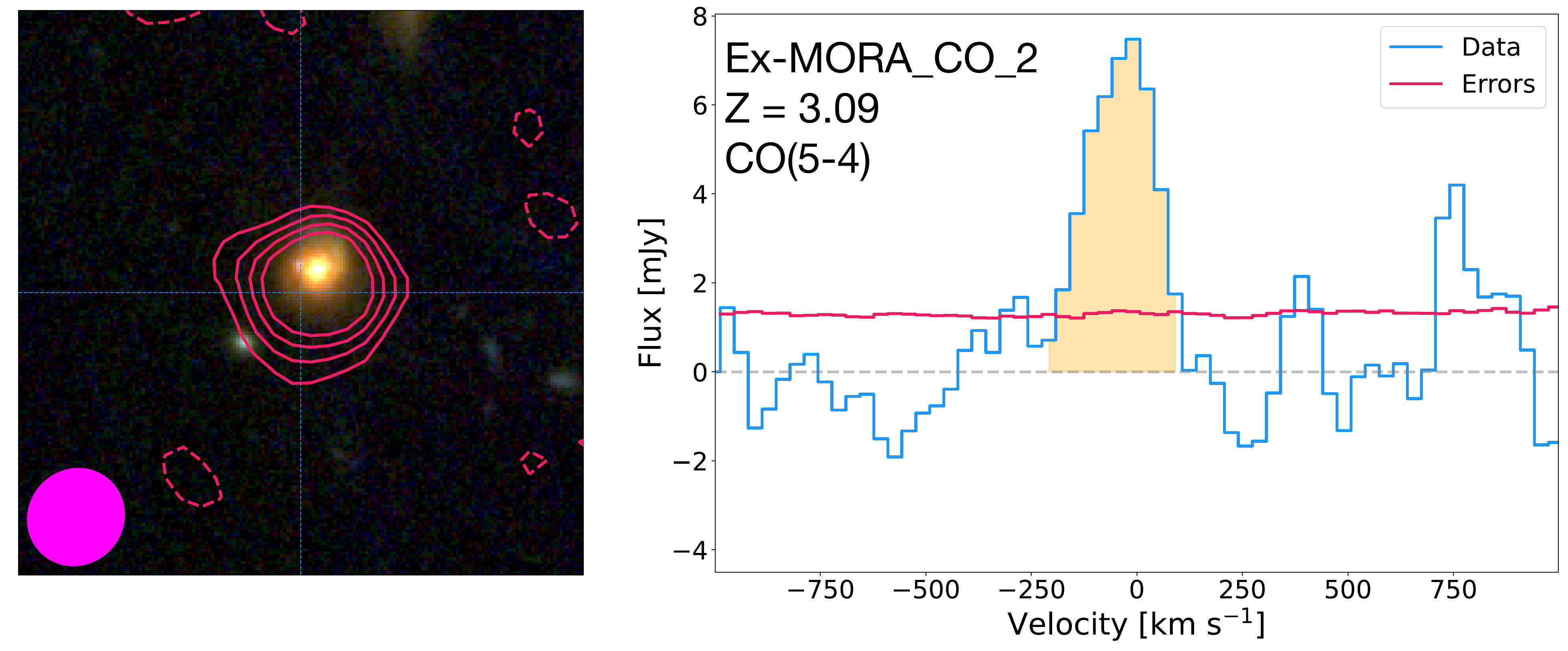}
\caption{Spectral and spatial characteristics of a representative unbiased detection from this work. The right panel shows the extracted emission-line spectrum (blue), the 1$\sigma$ errors (red), and the area collapsed for the integrated flux indicated by the orange shaded region. The left panel displays a three-color optical/NIR image of the source field (16"$\times$16"), with ALMA emission contours corresponding to  2$\sigma$, 3$\sigma$, 4$\sigma$, 5$\sigma$, where the corresponding negative contours are dashed lines. The solid magenta ellipse in the lower-left corner indicates the ALMA synthesized beam with a size of 1.58" $\times$ 1.50". }
\label{fig:blind_spectra}
\end{figure*}

Identification of the detected spectral lines, specifically determining which CO or \ci transition was observed, relied on the available auxiliary data. For sources with high-quality $z_{spec}$, identification was immediate, as the observed frequency could be directly converted to the rest-frame transition frequency. For the only source lacking a spectroscopic redshift, the redshift probability distributions (PDZs) provided by the COSMOS-Web DR1 and derived with LePhare SED fitting \citep{shuntovCOSMOS2025COSMOSWebGalaxy2025} was used. The most likely transition was evaluated by comparing the observed line frequency with the PDZ median. The line was confirmed when the resulting spectroscopic redshift ($z_{line}$) fell within the $2\sigma$ confidence interval of the photometric PDZ, thereby preventing mismatching. Using this unbiased detection approach, 18 high-fidelity detections were recovered. An example of an unbiased emission line is shown in Figure~\ref{fig:blind_spectra}, along with the emission line intensity of the galaxy over a composite color image.

\subsection{Targeted Search}
The unbiased search provided a census of high-luminosity line emitters but was limited by the high SNR threshold (SNR=6) required to maintain low false-detection rates across large survey volumes. To address this limitation, a targeted search was performed using the spectroscopic redshift compilation. Prior knowledge of the precise $z_{\text{spec}}$ enabled the probing of fainter emission at predefined frequencies and allowed for lower detection thresholds for known galaxies while maintaining catalog purity.

All galaxies in the spectroscopic catalog with coordinates within the Ex-MORA footprints were selected, ensuring that their expected CO or \ci transitions fell within the Band 4 range. At each source position, the galaxy spectrum was extracted, followed by a Gaussian fit centered on the expected observed frequency, allowing a 200 km s$^{-1}$ offset. If the fit exhibited a positive amplitude, the emission was integrated over a velocity range of $3\sigma$ of the profile to generate a zeroth-moment map. For negative peaks, no emission was assumed to be present. In other cases, the peak SNR was measured within a $1''$ radius centered on the optical/near-infrared position, accounting for possible offsets between stellar and gas centroids. This method is deteailed in Appendix \ref{app: noise}.

To determine the optimal SNR cut for the targeted sample and rigorously quantify the false-positive probability, a random-frequency control test was conducted. A sample was generated by assigning random redshifts to random points in the maps, and a distribution of simulated SNR values from noise fluctuations was constructed. These values were then compared with those from the spectroscopic sample. The objective was to establish a dynamic SNR for each cube, with the lowest SNR reaching 3.9. This threshold limits the final contamination (false-positive) rate to 1\%. This thresholding technique is essential for distinguishing accurate physical sources from inherent noise at known spectroscopic positions.

\begin{figure*}
\includegraphics[width=\textwidth]{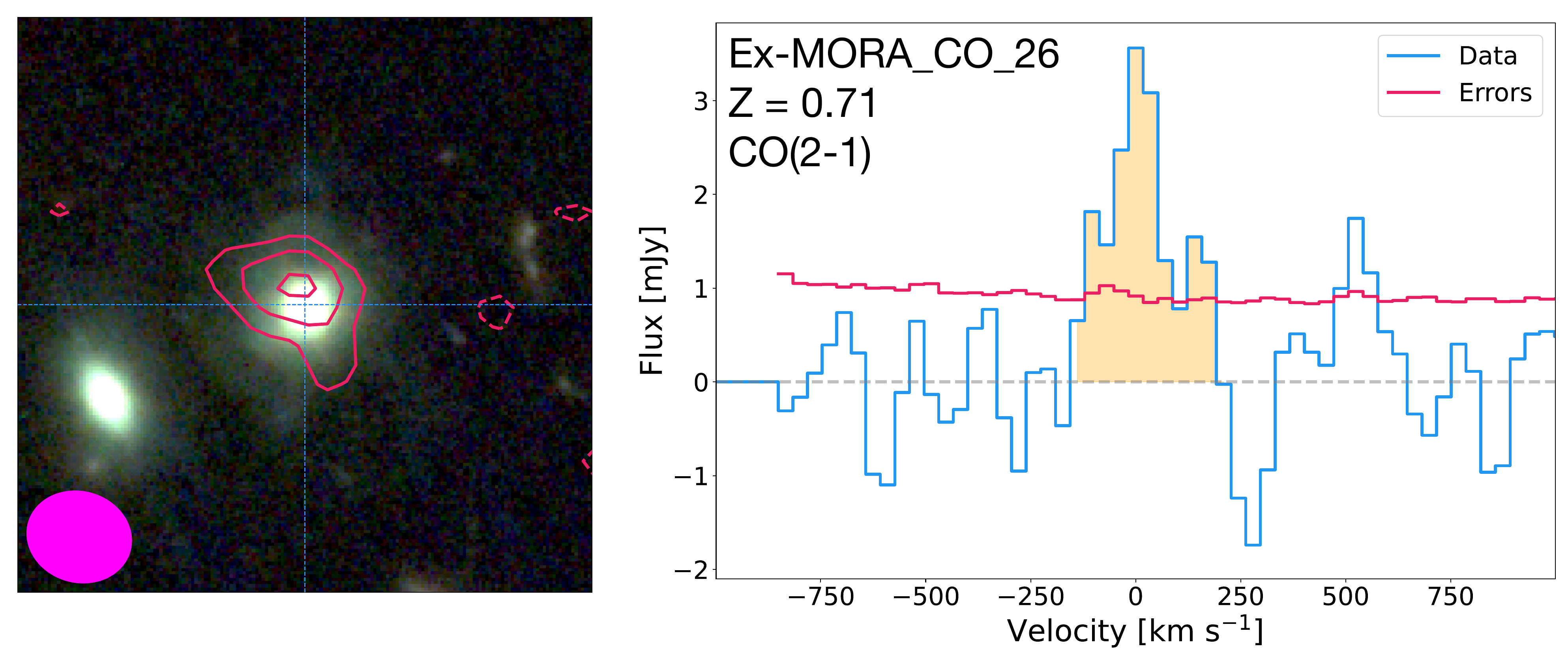}
    \caption{Same caption as Figure \ref{fig:blind_spectra} but for a targeted detection and a beam with a size of 1.68" $\times$ 1.47"}
    \label{fig:spec_spectra}
\end{figure*}

After cross-referencing this list with the unbiased results to eliminate duplicates, 34 additional galaxies were identified, nearly tripling the total sample size. This approach enabled the inclusion of lower-luminosity systems, which broadens our sampling of typical star-forming galaxies across the field. An example of a targeted emission line is shown in Figure~\ref{fig:spec_spectra}, along with the emission line intensity of the galaxy over a composite color image.

\subsection{Final catalog}

\begin{figure}
    \centering
    \includegraphics[width=1\linewidth]{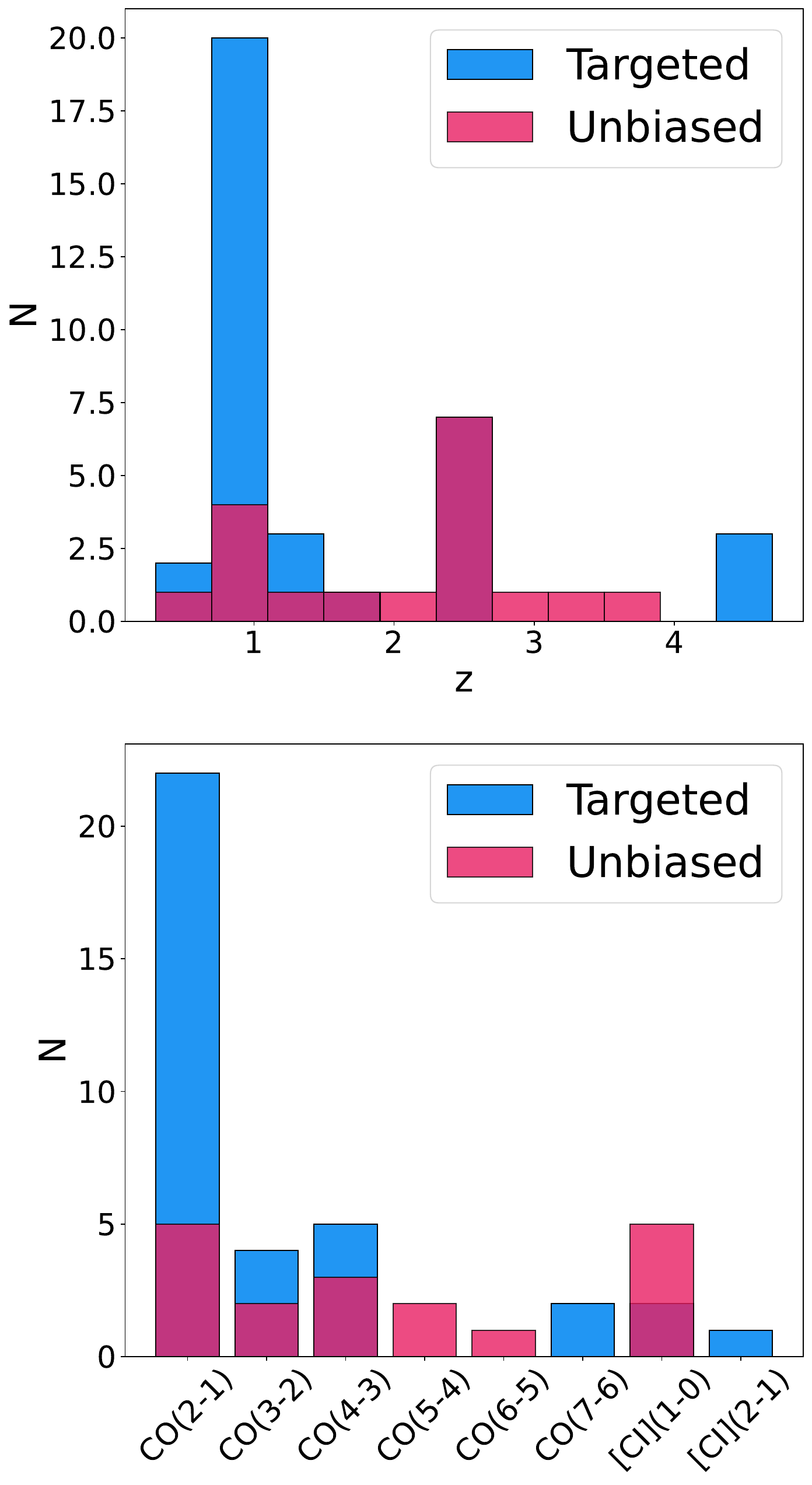}
    \caption{Distribution of spectroscopic redshifts and detected line species for the final galaxy sample. Both components of the graph (unbiased and targeted) begin at zero. Top panel: Redshift distribution ($z_{\text{spec}}$) of the 54 high-fidelity detections. Bottom panel: Distribution of detected transitions, comprising various rotational $J$ levels of CO and the fine-structure lines of \ci. The sample includes 18 unbiased detections and 36 targeted emissions identified via spectroscopic priors.}
    \label{fig:distribution}
\end{figure}

Figure~\ref{fig:distribution} presents the redshift and line distribution of the final sample. The final catalog comprises 54 high-fidelity emission detections, established through the combination of unbiased and targeted search strategies. Of these, 18 emissions were identified through the unbiased search algorithm, representing the most statistically robust emitters in the field, while the remaining 36 were recovered via spectroscopic targeting (two galaxies each showed two emission lines). As shown in the bottom panel of Figure~\ref{fig:distribution}, the sample consists exclusively of CO and \ci transitions, both of which are tracers of the cold ISM, and robust redshifts are available for all galaxies.

A primary feature of Figure~\ref{fig:distribution} is the prominent spike at $z \approx 0.73$, corresponding to detections of the CO(2-1) transition. This overdensity represents the spectroscopic signature of the "COSMOS Wall", a massive large-scale structure (LSS) first characterized by \cite{guzzoCosmicEvolutionSurvey2007}. Due to the large number of spectroscopic redshifts for galaxies at this redshift, most galaxies in this bin are targeted detections. An additional overdensity is observed at z$\sim$ 2.45, corresponding to a proto-supercluster first reported in \cite{cucciatiProgenyCosmicTitan2018}. The successful recovery of these structures demonstrates that the ALMA selection is sensitive enough to probe high-density environments, providing a laboratory to study how the local environment influences gas processing and star-formation efficiency. Interpretation of the redshift trends in Figure \ref{fig:distribution} and the detections within overdensities requires consideration of Figure \ref{fig:luminosities}. The observed redshift distribution is primarily determined by the selection function of ALMA Band 4. The clustering of detections largely results from bright emission lines entering the band, such as the peak of CO(2-1) detections at $z \approx 0.73$ and \ci at $z \approx 2.4$. Although the lower luminosity limit for CO(2-1) enables deeper probing of the galaxy population at lower redshifts, the localized spikes in the redshift distribution correspond to these line-band intersections. Consequently, these overdensities are detected because the observational setup is specifically sensitive to these redshift windows. The observed trends should follow the intrinsic galaxy population but be affected by different selections and sensitivities at different redshifts.

Figure~\ref{fig:detections} presents the spatial distribution of the detected sources alongside the continuum-detected sources from \cite{longExtendedMappingObscuration2024}. Among the 52 line-detected sources, 10 were also identified in the continuum by \cite{longExtendedMappingObscuration2024}. This limited overlap suggests that the line search probes a distinct, less obscured population compared to the continuum-selected sample, or that many continuum sources fall outside the targeted redshift and excitation ranges. The minimal intersection between the continuum- and line-detected samples further suggests that the line-detected sample is more strongly influenced by gas content than by excitation properties. Of the 10 overlapping sources, 6 emit in highly excited CO transitions (CO(4-3) to CO(7-6)), while only one emits CO(2-1). The remaining sources emit in \ci transitions: two galaxies emit (1-0), and one emits (2-1). The finding that most jointly detected galaxies exhibit both dust continuum and highly excited gas indicates a strong correlation between elevated, heavily obscured star formation rates (SFR) and the warmer, denser interstellar medium conditions characteristic of starburst (SB) galaxies. However, only two of these galaxies are classified as SB, while the remainder are main-sequence (MS) galaxies.

\section{SED fitting and galaxy properties}\label{sect:SED}

Key physical parameters, primarily stellar mass (M$_{*}$) and star formation rate (SFR), were estimated through spectral energy distribution (SED) fitting using the Code Investigating GALaxy Emission (\texttt{CIGALE v2025.1}; \citep{boquienCIGALEPythonCode2019}). \texttt{CIGALE} is based on the principle of energy balance, where ultraviolet (UV) and optical radiation from both young and old stellar populations is partially absorbed by interstellar dust. The absorbed energy is subsequently re-emitted in the infrared (IR) and submillimeter regimes \citep{boquienCIGALEPythonCode2019}.

Multi-wavelength photometry from the COSMOS-Web DR1 catalog \citep{shuntovCOSMOS2025COSMOSWebGalaxy2025} was utilized for our fits. Although the catalog provides CIGALE-derived properties, additional photometric data were incorporated into the fitting procedure to distinguish this analysis while maintaining consistency in results. For sources either absent from the catalog or with extended apertures not adequately captured by automated pipelines, manual aperture photometry was performed using the original NIRCam images. Apertures were tailored to each source's geometry to ensure accurate measurement of stellar light. Our analysis also included 2 mm continuum flux densities or upper limits from ALMA Band 4 observations and, when available, ALMA Band 7 archival data.

The model grid was constructed primarily following the methodology of \cite{arango-toroCOSMOSWebHistoryGalaxy2025} to ensure comparability with COSMOS-Web field studies. The specific parameters used in the Bayesian grid, including metallicity range, ionization parameters, and dust slopes, are summarized in Table~\ref{table: cigale}. The main differences in our approach were the inclusion of AGN feedback for the cases where AGN were present. To determine which galaxies were affected by AGN, we cross-matched our sample to the VLA-COSMOS 3 GHz  catalog from \cite{delvecchioVLACOSMOS3GHz2017}.

To minimize biases, a non-parametric approach was adopted by dividing the star formation history (SFH) into multiple independent time bins using the sfhNlevels module in \texttt{CIGALE}, following \cite{lejaHowMeasureGalaxy2019} and refined by \cite{cieslaGOODSALMA20Last2023}. In this framework, the SFR in each bin is treated as an unconstrained parameter, with bins linked by a continuity-burst prior. This prior, based on a Student's t-distribution, moderates transitions between bins by promoting smooth SFR evolution while permitting abrupt changes. The model was configured with seven discrete time bins to sample cosmic history from each galaxy's current redshift to the onset of star formation. While \cite{arango-toroCOSMOSWebHistoryGalaxy2025} employed ten bins, the default value of seven was used in this analysis. Table~\ref{table:properties} presents the key properties derived from the fitting.

\begin{table}
\centering
\caption{SED fitting parameters for CIGALE} \label{table: cigale}
\centering
\begin{tabular}{l l l }
\hline\hline
\multicolumn{3}{c}{\textbf{Non-parametric: sfhNlevels (1)}} \\
\hline
$age$ [Myr] & {[963; 13753]} & Age of the oldest stars in the   \\
&&galaxy in Myr; 100 in values \\
&&linearly sampled. \\
$N_{\text{SFH}}$ & 2000 & Number of SFHs drawn for each  \\
&&$age$ value. \\
\hline
\hline
\multicolumn{3}{c}{\textbf{Simple stellar population: BC03 (2)}} \\
\hline
IMF & Chabrier & Initial Mass Function \\
$Z$ & 0.008, 0.02 & Metallicity \\
\hline
\hline
\multicolumn{3}{c}{\textbf{Attenuation: dust\_att\_modified\_starburst (3)}} \\
\hline
$E(B-V)s$ & [0, 1.8] & Color excess; 10 values  \\
&& linearly sampled \\
\hline
\hline
\multicolumn{3}{c}{\textbf{Dust emission: dale2014 (4)}} \\
\hline
$\alpha$ & 2.0 & Far-IR slope \\

\hline
\hline
\multicolumn{3}{c}{\textbf{AGN: skirtor2016 (5)(6) (only for AGN)}} \\
\hline
fracAGN  & 0, 0.1, 0.2, 0.3,& AGN fraction \\
 & 0.4,0.5,0.6,0.7&\\

\hline
\hline
\end{tabular}
\tablefoot{ Listed are the parameters used to estimate the SED fits for all the galaxies in our sample, generally following the values used by \citet{arango-toroCOSMOSWebHistoryGalaxy2025}. \textbf{References}: (1): \citet{cieslaGOODSALMA20Last2023}, based on the methodology of \citet{lejaHowMeasureGalaxy2019}; (2): \cite{bruzualStellarPopulationSynthesis2003}; (3): \cite{calzettiDustContentOpacity2000}; (4): \cite{daleTwoparameterModelInfrared2014}; (5): \cite{stalevskiSKIRTORDatabaseModelled2012}; (6): \cite{stalevskiDustCoveringFactor2016}}
\end{table}

\begin{table*}
\centering
\caption{Galaxy properties} 
\begin{tabular}{l c c c c c c c}
\hline\hline
ID & RA & DEC & $z_{line}$ & Line & Log(SFR) & Log(M$_*$) & Log($M_{mol}$) \\
 & [deg] & [deg] & & & [M$_{\odot}$yr$^{-1}$]  & [M$_{\odot}$] & [M$_{\odot}$] \\
\hline
Ex-MORA\_CO\_1 & 150.12 & 2.534 & 3.3453(3) & CO(5-4) & 2.12$\pm$0.17 & 10.55$\pm$0.22 & 11.46$\pm$0.12 \\
Ex-MORA\_CO\_2 & 150.054 & 2.203 & 3.0908(1) & CO(5-4) & 2.28$\pm$0.29 & 11.32$\pm$0.06 & 11.01$\pm$0.12 \\
Ex-MORA\_CO\_3 & 150.105 & 2.313 & 2.29(1) & CO(4-3) & 2.67$\pm$0.13 & 10.82$\pm$0.12 & 10.77$\pm$0.13 \\
Ex-MORA\_CO\_4 & 150.086 & 2.298 & 1.2523(2) & CO(3-2) & 2.48$\pm$0.25 & 10.83$\pm$0.21 & 10.85$\pm$0.16 \\
Ex-MORA\_CO\_5 & 150.21 & 2.312 & 0.74729(7) & CO(2-1) & 2.27$\pm$0.24 & 10.66$\pm$0.14 & 10.44$\pm$0.08 \\
Ex-MORA\_CO\_6 & 150.147 & 2.337 & 0.7262(2) & CO(2-1) & 1.93$\pm$0.37 & 10.94$\pm$0.15 & 10.58$\pm$0.08 \\
\hline
\hline
\end{tabular}
\tablefoot{Extract of Table~\ref{table:properties2}. Derived values include Star Formation Rate (SFR) and stellar mass ($M_*$) from \texttt{CIGALE} SED modeling, and total molecular gas mass ($M_{mol}$). The number in parenthesis for $z_{line}$ corresponds to the error in the last decimal.}
\label{table:properties}
\end{table*}

\section{Molecular gas in our sample}\label{sec:6}

To quantify the total molecular gas available for star formation, the observed integrated line fluxes were converted into line luminosities, from which the molecular gas masses were derived. Table~\ref{table:properties2} presents the final $L’_{\text{line}}$ values for each galaxy. The conversion of these luminosities into $M_{\text{mol}}$ is highly dependent on the physical properties of the tracer. Given the fundamental differences in excitation conditions, optical depths, and chemical abundances between CO and \ci, the molecular gas mass was estimated separately for each species, incorporating their respective conversion factors ($\alpha_{\text{CO}}$ and $\alpha_{\text{\ci}}$). In both cases measurement errors were propagated correspondingly to obtain the error of the molecular gas masses. We note that the scatter of the relations were not considered in these errors.

\subsection{CO lines}

The molecular gas mass derived from the CO transition used the $\alpha_{\text{CO}}$ conversion factor. Because multiple J transitions were studied, the measured luminosities were first converted to CO(1-0) luminosities using the r$_{J1}$ parameters defined in Eq.~\ref{defr}.
\begin{equation} \label{defr}
r_{\rm{J1}} = \frac{{L’}_{\rm{CO(J-(J-1))}}}{{L’}_{\rm{CO(1-0)}}}
.\end{equation}

In this context, ${L’}_{\rm{CO(J-(J-1))}}$ denotes the luminosity of the corresponding J transition. According to \cite{boogaardALMASpectroscopicSurvey2020a}, galaxies with redshifts greater than 2 exhibit more highly excited CO SLEDs compared to those at lower redshifts. Due to this finding, we use different excitation parameters for different redshift intervals. The CO SLEDs of stacked galaxies were modeled in two redshift bins (\textit{z} < 2 and \textit{z} > 2) to determine average r$_{\rm{J1}}$ parameters. The CO(2-1) and CO(3-2) emitters in this sample have redshifts below 2, so the values estimated for this range were adopted: r$_{21}$ = 0.83 $\pm$ 0.12 and r$_{31}$ = 0.58 $\pm$ 0.10. For higher J transitions, the values for redshifts above 2 were used: r$_{41}$ = 0.76 $\pm$ 0.16, r$_{51}$ = 0.59 $\pm$ 0.14, and r$_{71}$ = 0.19 $\pm$ 0.09. For galaxies that showed AGN activity, the excitation factors were adjusted accordingly by using the \cite{kirkpatrickCOEmissionInfraredselected2019a} excitation values instead of \cite{boogaardALMASpectroscopicSurvey2020a}.

Following the determination of $L’_\mathrm{CO(1-0)}$ values, these were multiplied by $\alpha_{\text{CO}}$, which was derived using the metallicity-dependent relation from \cite{tacconiPHIBSSUnifiedScaling2018}. This approach enables direct comparison with their results. The resulting $\alpha_{\text{CO}}$ values ranged from 2.4 $\pm$ 0.01 to 3.5 $\pm$ 0.05 $\text{M}_\odot \ (\text{K} \ \text{km} \ \text{s}^{-1} \ \text{pc}^2)^{-1}$, which are below the typical value for Milky Way-like galaxies. This lower value is attributed to the predominance of high-mass galaxies in the sample, which generally exhibit higher metallicities and consequently lower $\alpha_{\text{CO}}$.

Due to the absence of direct metallicity measurements for the sample, the mass-metallicity relation from \cite{sandersMOSDEFSurveyEvolution2021} was employed. This study provides two relations for different redshift bins: above and below \textit{z} = 2. Stellar masses from Section~\ref{sect:SED} were used to calculate the final metallicity values, which enabled the determination of the $\alpha_{CO}$ parameter and its associated uncertainty for each galaxy.

\subsection{ \ci lines}

In addition to CO, the \ci fine-structure lines were used to estimate molecular gas mass. Similar to the CO-to-$H_2$ conversion factor, $\alpha_{\ci}$ is strongly dependent on the ISM gas-phase metallicity. To address this, the empirical relation from \cite{heintzDirectMeasurementLuminosity2020a} was adopted. As direct gas-phase metallicity measurements were not available for the sample, metallicity values were derived using the stellar mass-metallicity relation from \cite{sandersMOSDEFSurveyEvolution2021}, based on stellar masses from Section~\ref{sect:SED}. The resulting $\alpha_{\ci}$ values ranged from 10.37 $\pm$ 0.7 to 33.7 $\pm$ 4.

Both \ci(1-0) and \ci(2-1) transitions were detected in the observations. To ensure consistency in the molecular gas census, all neutral carbon measurements were standardized to \ci(1-0). For sources with only the higher-frequency \ci(2-1) line, an excitation correction factor of $R = 0.44 \pm 0.03$ was applied, following the average brightness temperature ratio reported by \cite{valentinoPropertiesInterstellarMedium2020} for high-redshift star-forming galaxies.

\section{Our galaxies in context}\label{sec:7}
To provide context for the sample, results were compared with two literature datasets: the PHIBBS compilation \citep{tacconiPHIBSSUnifiedScaling2018} and the A3COSMOS archival sample 2019 catalog \citep{liuAutomatedMiningALMA2019}. Sub-samples from these catalogs were selected to match the redshift range $0.5 < \textit{z} < 5$. The PHIBBS sample includes 833 galaxies with stellar masses $\log(M_*/M_\odot)$ between 9 and 12 and star formation rates $\log(\text{SFR}/M_\odot\,\text{yr}^{-1})$ between 0 and 3.5. The A3COSMOS sub-sample contains approximately 1200 galaxies with $\log(M_*/M_\odot)$ spanning 7 to 12 and $\log(\text{SFR})$ from -2 to 3.5. Both surveys estimate molecular gas masses using either CO or dust continuum emission. To address the broad cosmic time span, our results were divided into four redshift bins: $0.5 < \textit{z} \le 1$, $1 < \textit{z} \le 2$, $2 < \textit{z} \le 3$, and $3 < \textit{z} \le 5$.

\begin{figure*}[h]
    \centering
    \includegraphics[width=\linewidth]{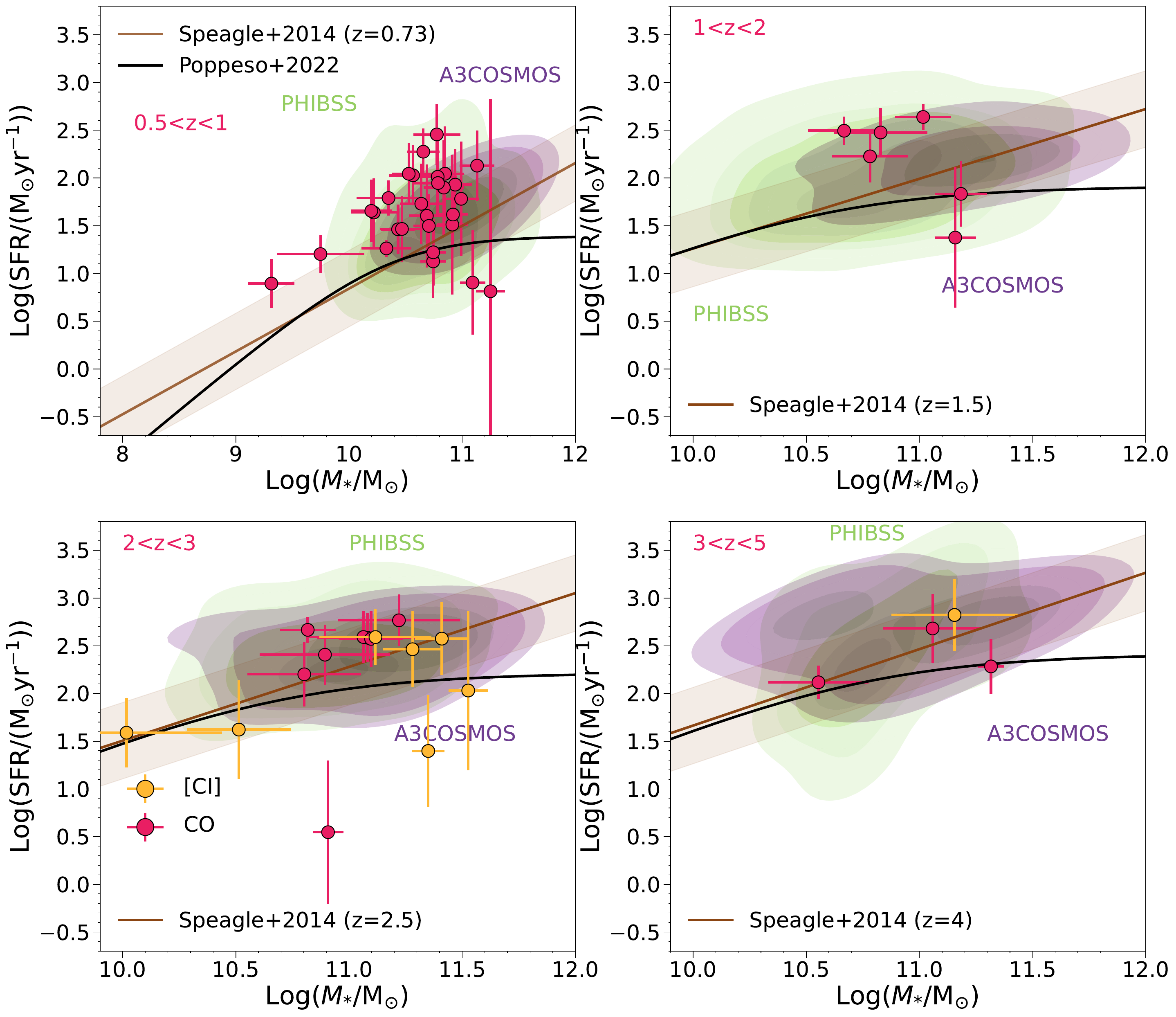}
    \caption{Galaxy MS for each redshift bin of our sample. Individual galaxies from our sample are shown as filled circles, color-coded by detected line species (red for CO; orange for \ci). The solid brown line denotes the main sequence (MS) relation from \citet{speagleHighlyConsistentFramework2014} at the median redshift of each bin, with the shaded region representing the intrinsic $\pm$ 0.4 dex scatter. On the other hand, the black line shows the corresponding \cite{popessoMainSequenceStarforming2023} MS for that redshift. For comparison, green and purple contours illustrate the density distributions of the PHIBSS \citep{tacconiPHIBSSUnifiedScaling2018} and A3COSMOS \citep{liuAutomatedMiningALMA2019} samples, respectively, within corresponding redshift intervals.}
    \label{fig: MS}
\end{figure*}

Figure~\ref{fig: MS} displays the sample on the $M_*$–SFR plane, compared to theMS relations from \citet{speagleHighlyConsistentFramework2014} (brown) and \cite{popessoMainSequenceStarforming2023} (black). In the lowest redshift bins, most galaxies were classified as SB ($0.4<\Delta MS$ dex) galaxies, followed by MS galaxies ($-0.4<\Delta MS<0.4$ dex), some green valley (GV; $-1.3< \Delta MS < -0.4$ dex) galaxies, and no passive systems  ($\Delta MS<-1.3$ dex). As \textit{z} increases, this changes, and MS galaxies dominate the sample. Overall, MS galaxies are the most prevalent, with 26 galaxies in this category, followed by 22 SB systems, 5 GV galaxies, and one passive galaxy. Using the \cite{popessoMainSequenceStarforming2023} MS as the classification criterion would increase the number of SB galaxies and decrease the number of GV and passive galaxies, particularly at higher stellar masses.

\begin{figure*}
    \centering
    \includegraphics[width=\linewidth]{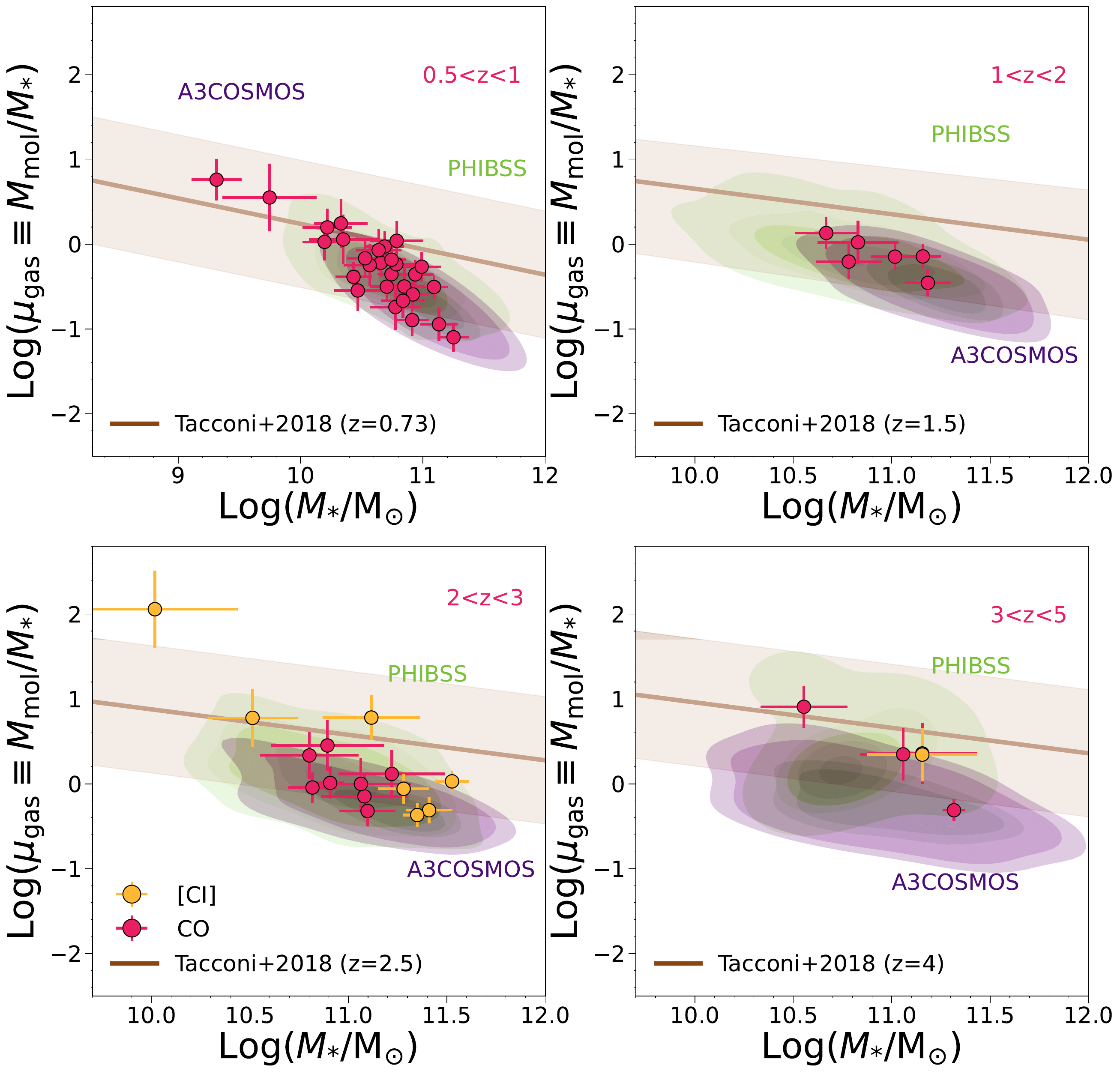}
    \caption{Molecular gas fraction in terms of stellar mass. Individual detections from our survey are represented by filled circles, color-coded by line species (red for CO; orange for \ci). The solid brown line denotes the empirical scaling relation extrapolated for MS from \citet{tacconiPHIBSSUnifiedScaling2018}, evaluated at the median redshift of each bin. For comparative context, green and purple contours show the density distributions of the PHIBSS and A3COSMOS samples, respectively.}
    \label{fig:mu}
\end{figure*}

The molecular gas to stellar mass ratio ($\mu_{\text{gas}} = M_{mol}/M_*$) measures the abundance of molecular gas relative to stellar mass. Figure~\ref{fig:mu} compares $\mu_{\text{gas}}$ values to the main sequence galaxy scaling relations extrapolated by \citet{tacconiPHIBSSUnifiedScaling2018}. Most galaxies follow the expected trend of increasing gas fraction with redshift, reflecting higher gas content at earlier cosmic times. Some galaxies fall below this trend, possibly indicating gas deficiency likely due to recent quenching or environmental stripping. In contrast, galaxies with elevated molecular gas content may have recently accreted fresh gas.

\begin{figure}
    \centering
    \includegraphics[width=\linewidth]{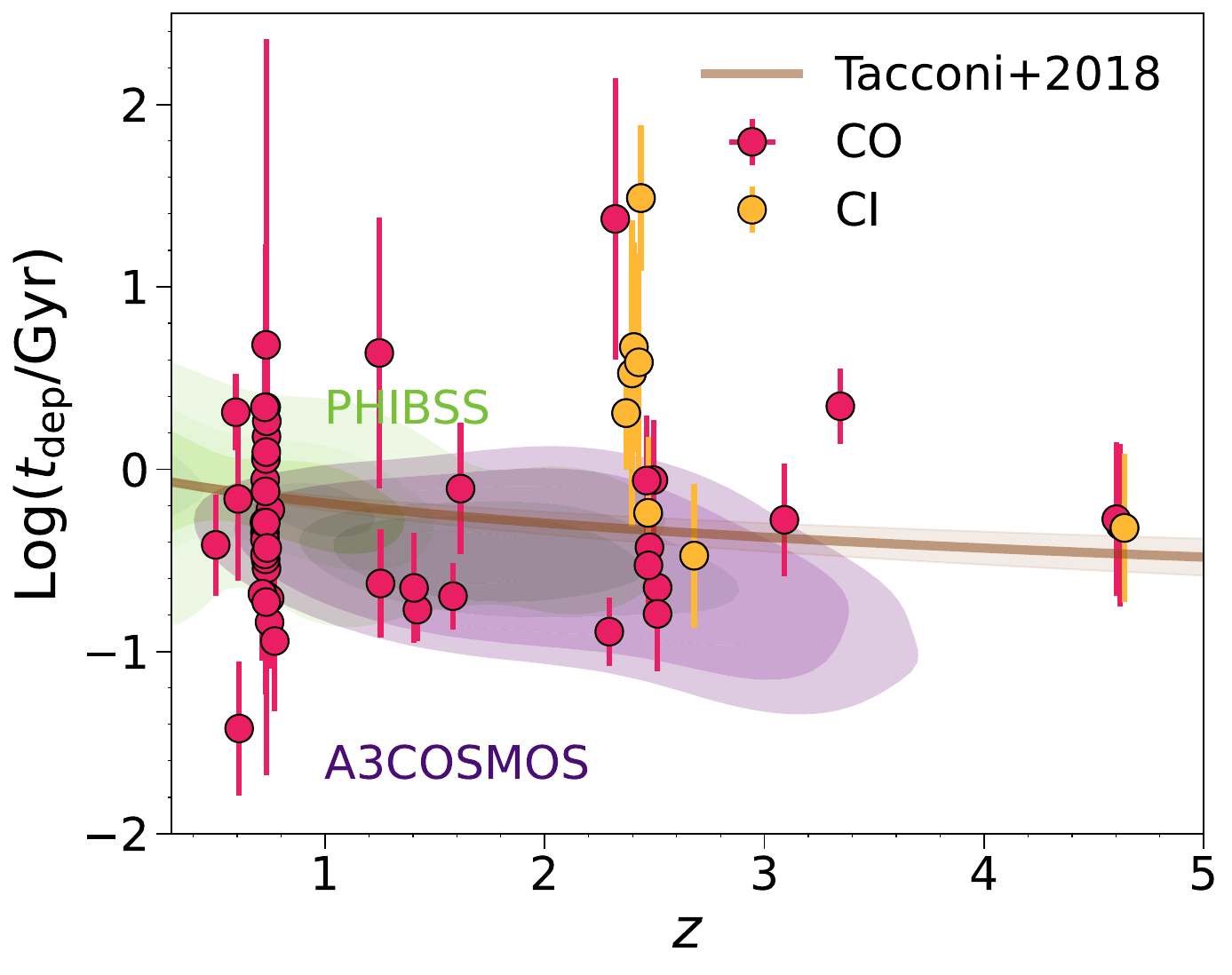}
    \caption{Depletion time in terms of redshift for four redshift bins. Individual detections from our survey are represented by filled circles, color-coded by line species (red for CO; orange for \ci). The solid brown line denotes the empirical scaling relation extrapolated for main-sequence galaxies from \citet{tacconiPHIBSSUnifiedScaling2018}, evaluated at the median redshift of each bin. For comparative context, green and purple contours show the density distributions of the PHIBSS and A3COSMOS samples, respectively.}
    \label{fig: tdep}
\end{figure}

The molecular gas depletion timescale ($t_{dep}$), defined as the inverse of the star formation efficiency (SFE), represents the duration over which galaxies can maintain their current star formation rates. High $t_{dep}$ values indicate slower gas consumption. As illustrated in Figure~\ref{fig: tdep}, many galaxies in the sample have $t_{dep}$ values exceeding those predicted by MS scaling relations, implying less efficient star formation compared to typical MS galaxies. In contrast, a significant fraction of galaxies exhibit short depletion times, characteristic of starburst activity in which gas is rapidly converted into stars. The observed variation within similar redshift and mass ranges suggests that gas-to-star conversion is more strongly governed by ISM conditions than by redshift or mass alone, assuming no recent gas infall.

\section{Discussion}\label{sec:8}
\subsection{Galaxies in our sample}

\begin{figure*}
    \centering
    \includegraphics[width=\linewidth]{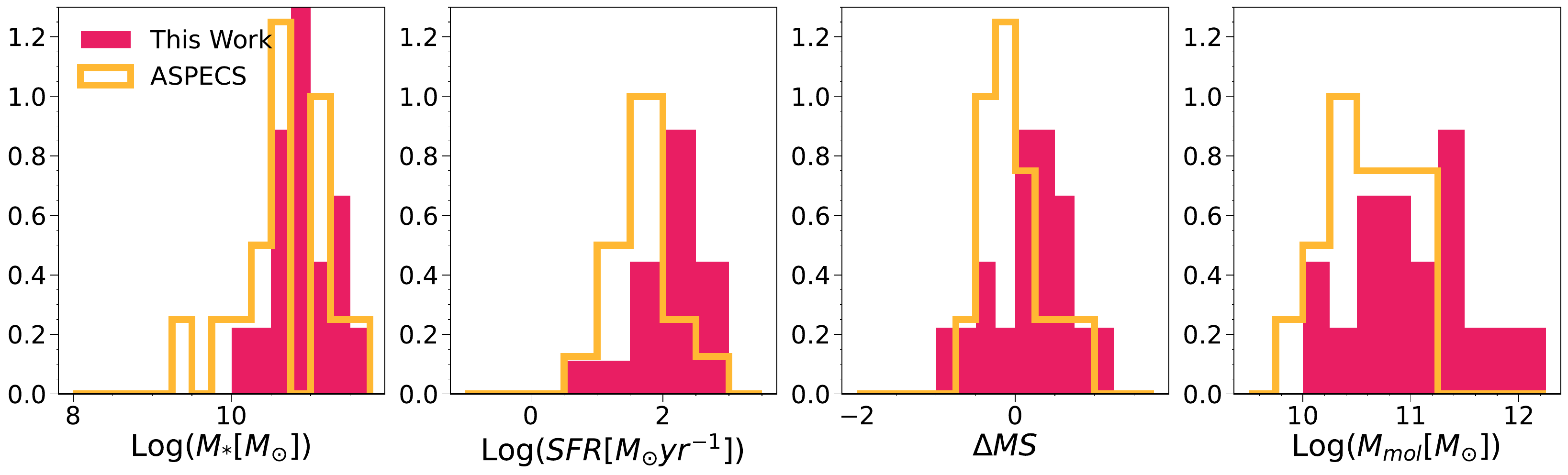}
    \caption{Normalized distribution of physical properties for the galaxy sample. The panels show stellar mass ($\log M_*$), star formation rate ($\log \text{SFR}$), distance from the main sequence ($\Delta MS$) , and molecular gas mass ($\log M_{mol}$). The orange line indicates the \cite{aravenaALMASpectroscopicSurvey2019} dataset, while the red represents our work.}
    \label{fig:properties2}
\end{figure*}

Because the selection technique targets molecular gas, it is essential to categorize the nature of the galaxies identified by this method. Selection based on molecular gas emission identifies galaxies by their fuel reservoirs instead of their ultraviolet luminosity. Figure~\ref{fig:properties2} presents the global distributions of physical properties for only the unbiased sample and compares them with those of the ASPECS sample \cite{aravenaALMASpectroscopicSurvey2019}, illustrating the diversity of galaxies identified in the Ex-MORA field.

A primary diagnostic of the sample's evolutionary state is the distance from the main sequence (MS), denoted $\Delta{MS}$, defined as the ratio between the specific star formation rate (sSFR) and the sSFR of the MS at a given mass ($\Delta{MS}$ = $sSFR/sSFR(MS)$). The galaxies in this sample exhibit $\Delta{MS}$ values ranging from $-1.62$ to $1.22$ dex, with 22 galaxies above the MS scatter, 6 below, and 26 within it. These results indicate that the method identifies galaxies at various evolutionary stages, with a predominance of MS galaxies. The identification of 5 GV galaxies and 1 passive galaxy, which are often underrepresented in high-redshift studies due to survey selection, further demonstrates that gas-based selection can identify galaxies undergoing quenching that still retain molecular gas.

The sample exhibits a high stellar-mass distribution, with a median value of $\log(M_*/M_\odot) = 10.83$. Most identified systems are high-mass, indicating that this technique preferentially selects more massive galaxies. The sample's star formation rate (SFR) is diverse, ranging from $3.5 \, M_{\odot}\text{yr}^{-1}$ to over $650 \, M_{\odot}\text{yr}^{-1}$, with a median value of $\log$(SFR [M$_{\odot}$yr$^{-1}$]) = 1.98. The focus on molecular gas tracers ensures that every galaxy in the sample possesses substantial gas mass, with most exceeding $10^{10} M_{\odot}$. The median molecular gas mass is $\log$(M$_{mol}/M_{\odot}$) = 10.69, with values reaching up to $\log$(M$_{mol}/M_{\odot}$) = 12.

The unbiased dataset was compared with ASPECS \citep{decarliALMASpectroscopicSurvey2016,aravenaALMASpectroscopicSurvey2019,gonzalez-lopezAtacamaLargeMillimeter2019}, a previous unbiased survey that covers a smaller area, as shown in Figure~\ref{fig:properties2}. In terms of stellar mass (M$_*$), both surveys recover galaxies over similar mass ranges, as indicated by the failure to reject the null hypothesis in the Kolmogorov-Smirnov (KS) test. For the SFR, our sample is slightly skewed toward higher values than ASPECS, suggesting that this selection recovers more actively star-forming systems; this is supported by the KS test, which indicates a p-value under 0.05. The distribution of $\Delta$MS is similar to that of ASPECS, but with higher values; however, the KS test finds these differences are not significant. Regarding molecular gas mass (M$_{mol}$), our sample generally traces galaxies with larger gas reservoirs. Therefore, the selection technique recovers galaxies with higher molecular gas masses; however, the difference is not statistically significant according to the KS test.

\subsection{Molecular gas depending on galaxy evolution state}

\begin{figure}
    \centering
    \includegraphics[width=\linewidth]{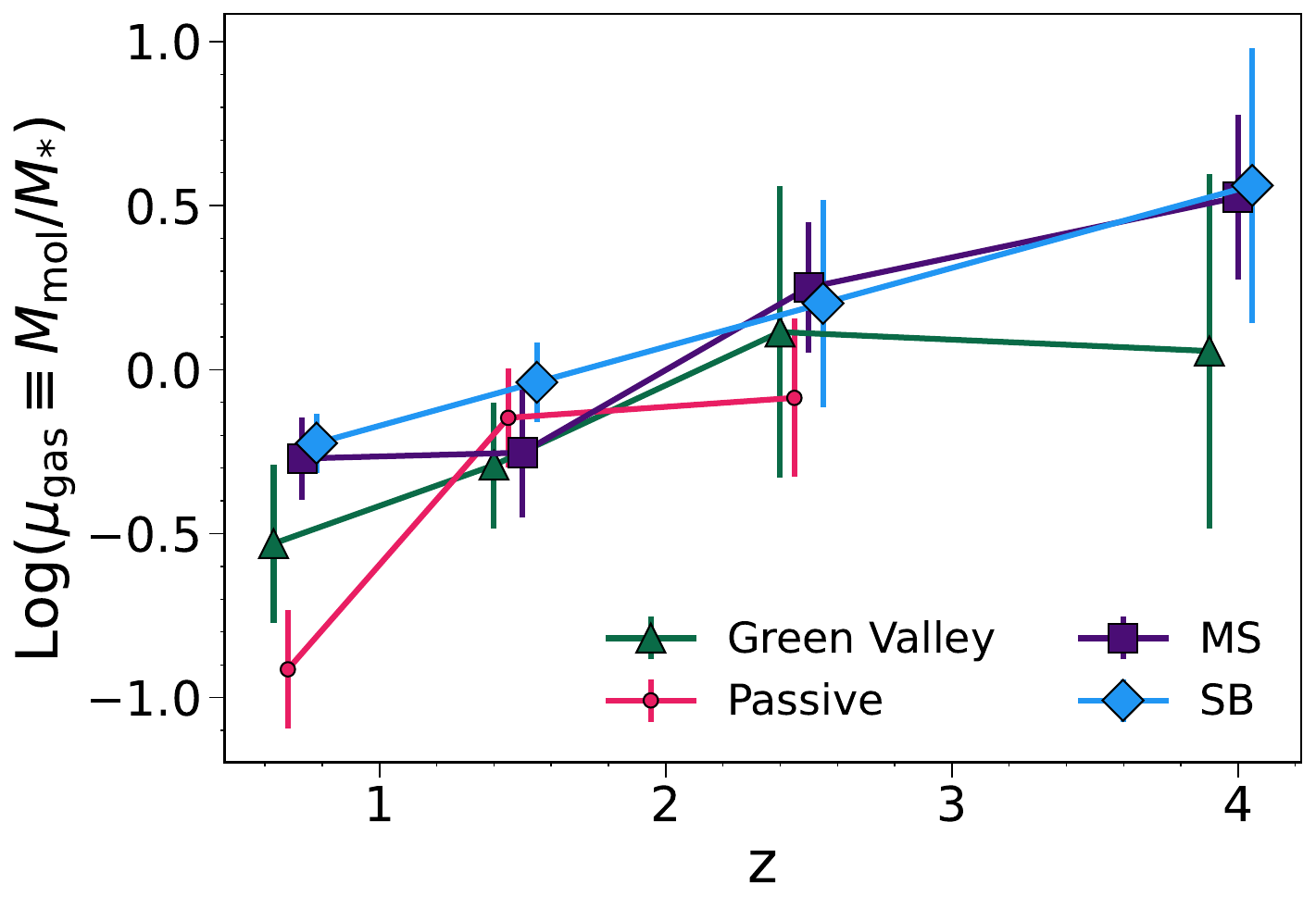}
    \caption{Evolution of average molecular gas fractions ($\mu_{\text{gas}}$) as a function of redshift ($z$). Galaxies are categorized by their star-forming state: Green Valley (green triangles), Main Sequence (MS; purple squares), Starburst (SB; blue diamonds), and passive galaxies (red circles), with $1\sigma$ uncertainties indicated by vertical bars.}
    \label{fig:classification_mu}
\end{figure}

To better characterize gas reservoirs in the sample, galaxies were separated into passive, GV, MS, and SB types. Galaxies with $-0.4<\Delta MS <0.4$ were categorized as MS, those with higher values as SB, those with $-1.3<\Delta MS <-0.4$ as GV, and the remainder as passive. The redshift evolution of average $\mu_{\text{gas}}$ is presented in Figure~\ref{fig:classification_mu}. To account for uncertainties in $\Delta\text{MS}$ and $\mu_{\text{gas}}$ for each galaxy, a Monte Carlo resampling approach was employed. In each of 10,000 iterations, the properties of each galaxy were randomly drawn from a Gaussian distribution centered on its observed value with a standard deviation corresponding to its measurement error. The final data points in Figure~\ref{fig:classification_mu} represent the median values across all iterations, with error bars denoting the $1\sigma$ standard deviation of the simulated distribution.

Despite the non-negligible error bars, we can see that MS and SB systems' $\mu_{\text{gas}}$ increases with redshift, consistent with previous scaling relations, such as those from \citet{tacconiPHIBSSUnifiedScaling2018}. There is a trend for GV and passive galaxies to have lower gas fractions than star-forming galaxies, most noticeable in the lower-redshift bin.

\subsection{Galaxy Kinematics}\label{sect:kine}
To further investigate galaxy behavior, kinematic analysis was conducted where feasible. First- and second-moment maps were derived using a multi-stage imaging pipeline designed to optimize signal-to-noise ratio while preserving physical structures. The data cube was initially smoothed spatially and spectrally using Gaussian convolution to reduce high-frequency noise and better match the resolution of emission features. After smoothing, a $3\sigma$ masking threshold was applied to isolate statistically significant signals from background noise, thereby preventing low-level artifacts from biasing the results. The processed cubes were then integrated using the CASA \texttt{immoments} task. Analysis of the first moment and composite color images enabled classification of galaxies into three categories: disks, mergers, and undefined. Disk galaxies exhibited a clear velocity gradient in the first moment, as shown in Figure \ref{fig:moment1} in Appendix \ref{app:kine}, with 16 galaxies meeting this criterion. Mergers were identified by a velocity gradient in conjunction with the composite color image, as illustrated in Figure \ref{fig:moment2} in Appendix \ref{app:kine}, with two galaxies classified as mergers. When classification was ambiguous, galaxies were designated as undetermined; an example is shown in Figure \ref{fig:moment3} in Appendix \ref{app:kine}. Most galaxies were classified as disks based on their kinematics, consistent with the visual inspection in Appendix \ref{sect:morpho}. It is important to note that extended disks are intrinsically easier to classify than compact or merging systems in this dataset. Extended disks produce clear velocity gradients in first-moment maps even at modest SNR and spatial resolutions. In contrast, compact systems are often unresolved or marginally resolved at the larger beam sizes of Ex-MORA, resulting in their kinematic structure being beam-smeared into an ambiguous signal. Mergers are similarly challenging: at these redshifts and resolutions, close pairs or post-coalescence systems can mimic disk-like velocity gradients, leading to potential misclassifications. Therefore, the disk fraction reported here likely represents an upper limit, while the merger fraction represents a lower limit.

\subsection{Effect of the overdensities}\label{sect:overdensities}

\begin{figure} 
    \centering
    \includegraphics[width=\linewidth]{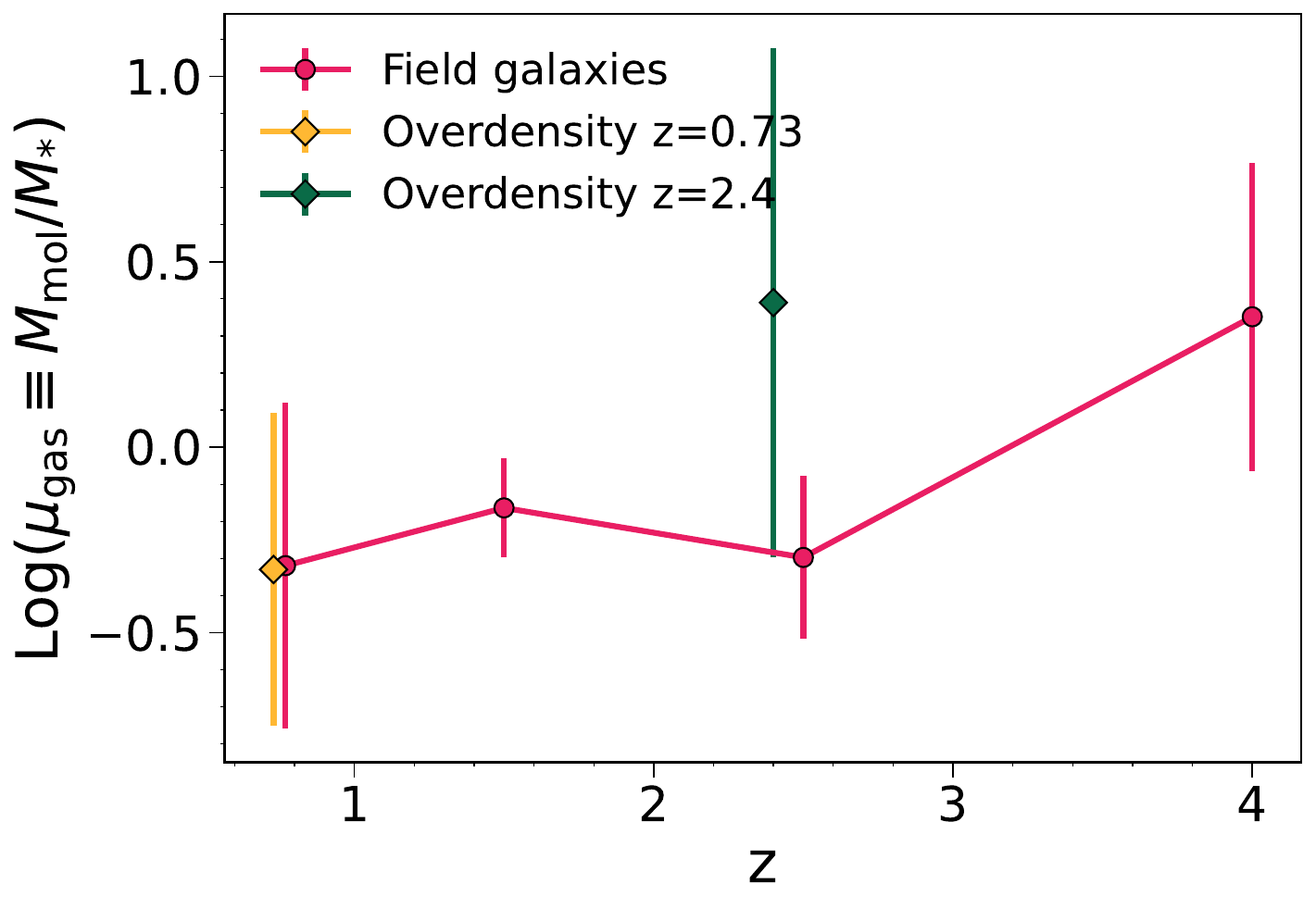}
    \caption{Evolution of average molecular gas fractions ($\mu_{\text{gas}}$) as a function of redshift ($z$) with $1\sigma$ uncertainties indicated by vertical bars. Galaxies are separated into three categories: the main sample (red), the overdensity at $z\approx 0.73$ (orange), and the overdensity at $z\approx 2.4$ (green).}
    \label{fig:overdensity}
\end{figure}

A portion of the sample is located within overdensities at $z\approx 0.73$ or $z\approx 2.4$. It is therefore necessary to assess whether the presence of these structures affects the main results. To examine this effect, the sample was divided into three categories based on their membership in the overdensity. The average $\mu_{\text{gas}}$ of galaxies not located in any overdensity is shown in red in Figure \ref{fig:overdensity} and is referred to as the field-galaxy population. This designation is nominal, as specific environmental effects have not been studied in detail. The average for galaxies in the overdensity at $z\approx 0.73$ is shown in orange, while that for the overdensity at $z\approx 2.4$ is shown in green. As shown in Figure \ref{fig:overdensity}, the values for field galaxies and the overdensity at $z\approx 0.73$ are similar, indicating that this overdensity does not significantly affect the results. This consistency may be due to the variety of environments within the COSMOS wall, ranging from dense clusters to field galaxies, which is similar to that expected for galaxies in the same redshift bin. 

In contrast, at $z\approx 2.4$, field galaxies have significantly lower gas fractions than those in the overdensity. This suggests that the sample is biased by the inclusion of this gas-rich overdensity. At this redshift, only four galaxies do not belong to the overdensity, all with gas fractions under 1. In comparison, the 11 galaxies in the overdensity have a median $\mu_{\text{gas}}$ above 1, with one galaxy hosting an extremely large gas reservoir. According to the study of \cite{gururajanGasPropertiesFunction2025}, the average $\mu_{\text{gas}}$ for galaxies in the protocluster should be around 0.83, yet even when excluding the extremely gas-rich source, the present sample yields an average of $\mu_{\text{gas}}=1.6$. This difference may result from our sample design, which identifies gas-rich galaxies and thereby misses galaxies with lower gas content, as included in most studies. This protocluster contains the most gas-rich galaxies within this redshift bin for our sample, which biases the results for this period. Notably, three galaxies are below the MS in the overdensity, which may represent systems in which intense star formation at cosmic noon heated the galaxies and abruptly halted star formation.

\section{Summary and Conclusions}\label{sec:9}

A molecular gas analysis was conducted for 52 galaxies (54 emission lines) within the COSMOS-Web field. ALMA Band 4 data from the Ex-MORA survey provided extensive spatial and spectral coverage over a combined area of 577 arcmin$^2$. Using the previously collapsed spectral axis, a comprehensive line search was performed. The methodology included both an unbiased line search to identify high-signal-to-noise emitters and a targeted spectroscopic search based on known redshifts. This dual approach enabled the detection of CO and \ci transitions across a broad redshift range ($0.5 < \textit{z} < 5$). Both emissions served as molecular gas tracers, facilitating the characterization of molecular gas properties within the sample. Physical parameters of the galaxies were estimated using non-parametric star formation history (SFH) modeling with \texttt{CIGALE}, which enabled flexible reconstruction of galaxy evolution and their SFHs.

Analysis of the molecular gas-selected galaxies reveals a population dominated by MS galaxies (26 galaxies), with additional representation from SB, passive, and GV sources. The sample is characterized by high average stellar mass and substantial molecular gas masses, typically exceeding $M_{mol}=10^{10}M_\odot$. Compared to ASPECS, this selection generally identifies galaxies with slightly higher molecular gas masses and SFR, suggesting a preference for more massive gas reservoirs than in traditional pencil-beam spectral scans.

The redshift evolution of the molecular gas fraction ($\mu_{\text{gas}}$) was investigated by categorizing the sample into different evolutionary stages based on their distance from the MS ($\Delta\text{MS}$). Across all active galaxy populations, $\mu_{\text{gas}}$ increases toward higher redshifts, reproducing established literature scaling relations and confirming that the cosmic evolution of star formation is fundamentally driven by the availability of larger gas reservoirs. SB and MS galaxies tend to have similar $\mu_{\text{gas}}$, with GV and passive systems presenting slightly lower values. It is important to note that, given the large error bars, this difference in the intermediate redshifts might not be significant.

In systems with extended structures, kinematical analyses were possible. These systems generally exhibit a clear velocity gradient, which, when combined with morphological analysis, indicates the presence of rotating disks. The prevalence of disks may be explained by the fact that extended disks are more easily detectable than compact, irregular galaxies within the survey's synthesized beam.

The impact of environmental overdensities at $z \approx 0.73$ and $z \approx 2.4$ on the global results was also evaluated. The overdensity at $z \approx 0.73$ exhibits average molecular gas fractions ($\mu_{\text{gas}}$) consistent with the field-galaxy population, likely reflecting the diverse environmental mix of the COSMOS wall. In contrast, the protocluster at $z \approx 2.4$ significantly biases the results, as galaxies within this overdensity display elevated gas fractions (generally $\mu_{\text{gas}} > 1$) compared to their field counterparts. The presence of GV systems within this protocluster suggests the posibility of a population in which intense star formation during Cosmic Noon may have heated the gas reservoir, prematurely halting star formation. These findings could indicate that while the lower-redshift overdensity does not significantly affect the results, the cosmic noon galaxies are biased toward gas-rich structures.

Although the majority of the sample consists of MS galaxies, the dual-approach methodology also enabled the detection of GV and passive galaxies, which provide valuable insights into galaxy evolution. Thus, the study successfully recovers galaxies at diverse evolutionary stages that contain molecular gas, making the main advantage over previous pencil-beam spectral scans the ability to recover large structures and higher gas masses.

\textit{Acknowledgments}

We thank the anonymous referee for their time and effort in reviewing this manuscript. We thank L. Ciesla for providing the code for the nonparametric SFR and K. Harrington for helping us obtain auxiliary data. We also acknowledge and thank the financial support from MINGAL (ANID - MILENIO - NCN2024\_112). We thank IA-PUC for their support. We thank Ian Smail for his suggestion. MA is supported by FONDECYT grant number 1252054 and gratefully acknowledges support from: ANID Basal Project FB210003,  ANID MILENIO NCN2024\_112 and ANID + Vinculaci\'on Internacional + FOVI250261. ET acknowledges support from the ANID CATA-BASAL program FB210003 and FONDECYT Regular 1241005 and 1250821.

\bibliographystyle{aa_url}
\bibliography{references}

\newpage
\begin{appendix} 
 \onecolumn

\section{Source Finding and Noise Characterization} \label{app: noise}
\subsection{The \texttt{Lineseeker} code}
The unbiased search employed the \texttt{Lineseeker} algorithm as described in \cite{gonzalez-lopezAtacamaLargeMillimeter2019} as used for the ALMA Large Program ASPECS. This approach initiates by convolving the spectral cubes with a set of Gaussian kernels spanning spectral widths from 50 to 500 km s$^{-1}$, thereby capturing a wide range of galaxy kinematics. Since no spatial convolutions are used, \texttt{Lineseeker} assumes that most of the emission falls within a single synthesized beam, which is a valid assumption for the beam of the EX-MORA observations. 

Data cubes created from ALMA observations are expected to have Gaussian noise background if no calibration problems are present. In this case, the Gaussian convolution in the spectral axis will produce a convolved cube also with a Gaussian nose. Since the spectral Gaussian convolution modifies the noise distribution (makes it have a larger variance), high-significance features are identified on a channel-by-channel basis. The initial noise level for each channel is estimated using the standard deviation of all corresponding voxels. To avoid artificial inflation of the noise estimate by bright emission lines, the calculation is repeated using only voxels with absolute values less than five times the initial noise estimate. The SNR for each voxel is then determined as the measured flux density in the collapsed channels divided by this empirically refined noise value.

After applying a SNR cut for the convolved data cube, source candidates are grouped using the Density-Based Spatial Clustering of Applications with Noise (DBSCAN) algorithm, with each final candidate assigned the maximum SNR observed across all convolutions. To address the non-independence of the applied spectral kernels, \texttt{Lineseeker} assesses detection fidelity by generating a conservative \texttt{PPessimistic} parameter which corrects for the Look-elsewhere effect assuming independent searches. A \texttt{PPessimistic} value of 0 indicates a highly reliable detection with no corresponding negative features at that significance  \citep{gonzalez-lopezAtacamaLargeMillimeter2019}.

\subsection{Targeted Search and Dynamic SNR Thresholding}
To complement the unbiased search and recover fainter, lower-luminosity systems, a targeted search was conducted using the spectroscopic redshift compilation from \cite{khostovanCOSMOSSpectroscopicRedshift2025}. For each cataloged galaxy with expected CO or [CI] transitions within Band 4, a localized spectrum was extracted. A Gaussian fit was performed centered on the expected observed frequency; if the fit was positive, the emission was integrated over a $3\sigma$ velocity range to generate a zeroth-moment map. The peak SNR was measured within a 1 arcsec radius to account for potential offsets between stellar and gas centroids. As only a single point was considered for this peak value, the image's root-mean-square (RMS) was used as the noise estimate.

Since our extraction method will prioritize positive signals, the SNR distribution will be skewed towards positive values and will not follow a Gaussian distribution. To rigorously define the detection threshold in these cases, a random-frequency control test was conducted. Random frequencies and spatial coordinates within the map were assigned to extract a simulated peak SNR distribution generated solely by the data’s inherent noise. By comparing the targeted extractions against this empirical noise baseline, a dynamic SNR threshold was established that strictly limits the false-positive contamination rate to below 1\%. This empirical thresholding effectively distinguished true physical sources from localized noise artifacts, supporting the addition of 39 galaxies to the final catalog.

\section{Morphological classification}

\begin{figure*}[htbp]
     \centering
     \begin{subfigure}[b]{0.24\textwidth}
         \centering
         \includegraphics[width=\textwidth]{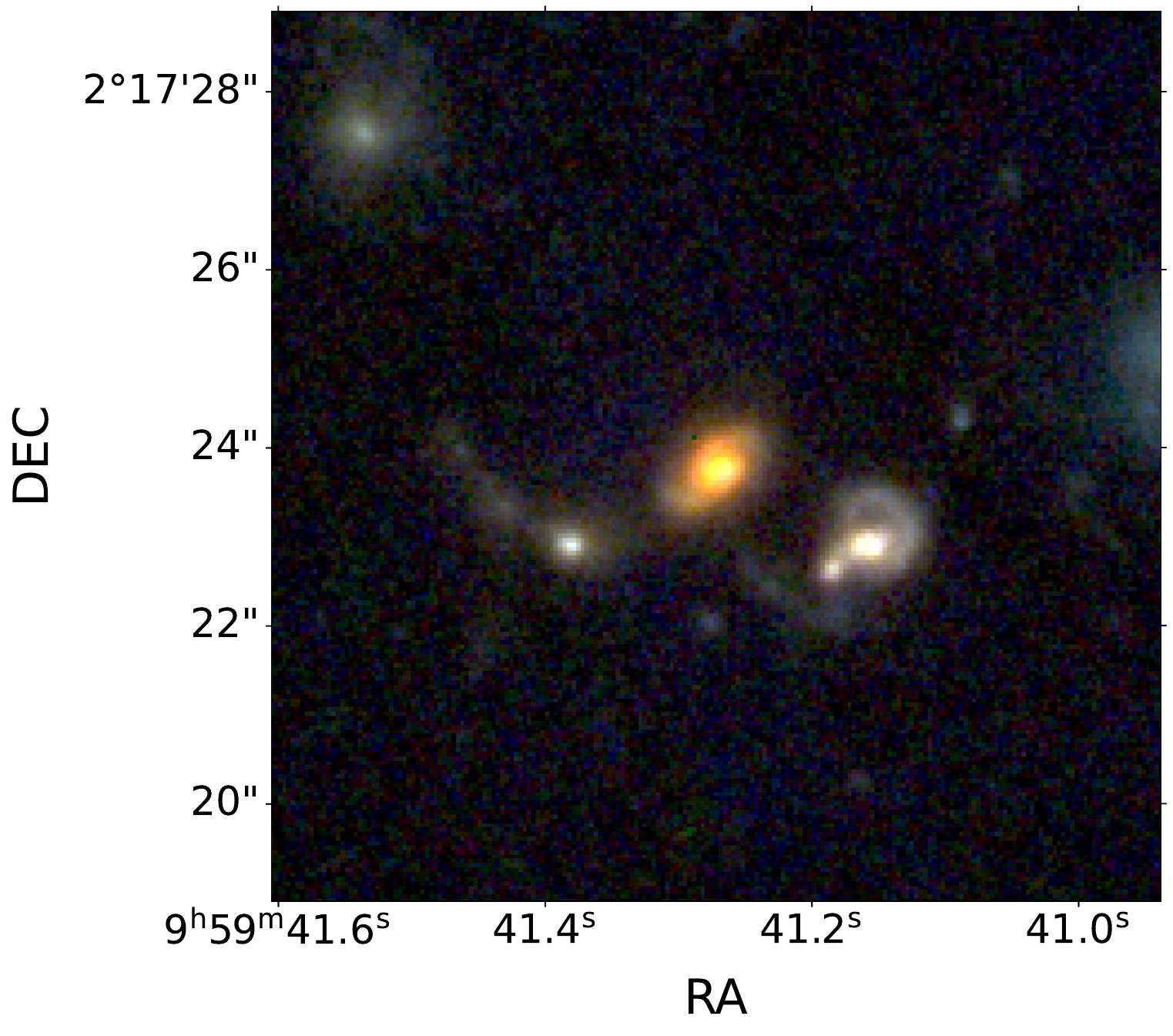}
     \end{subfigure}
     \hfill 
     \begin{subfigure}[b]{0.24\textwidth}
         \centering
         \includegraphics[width=\textwidth]{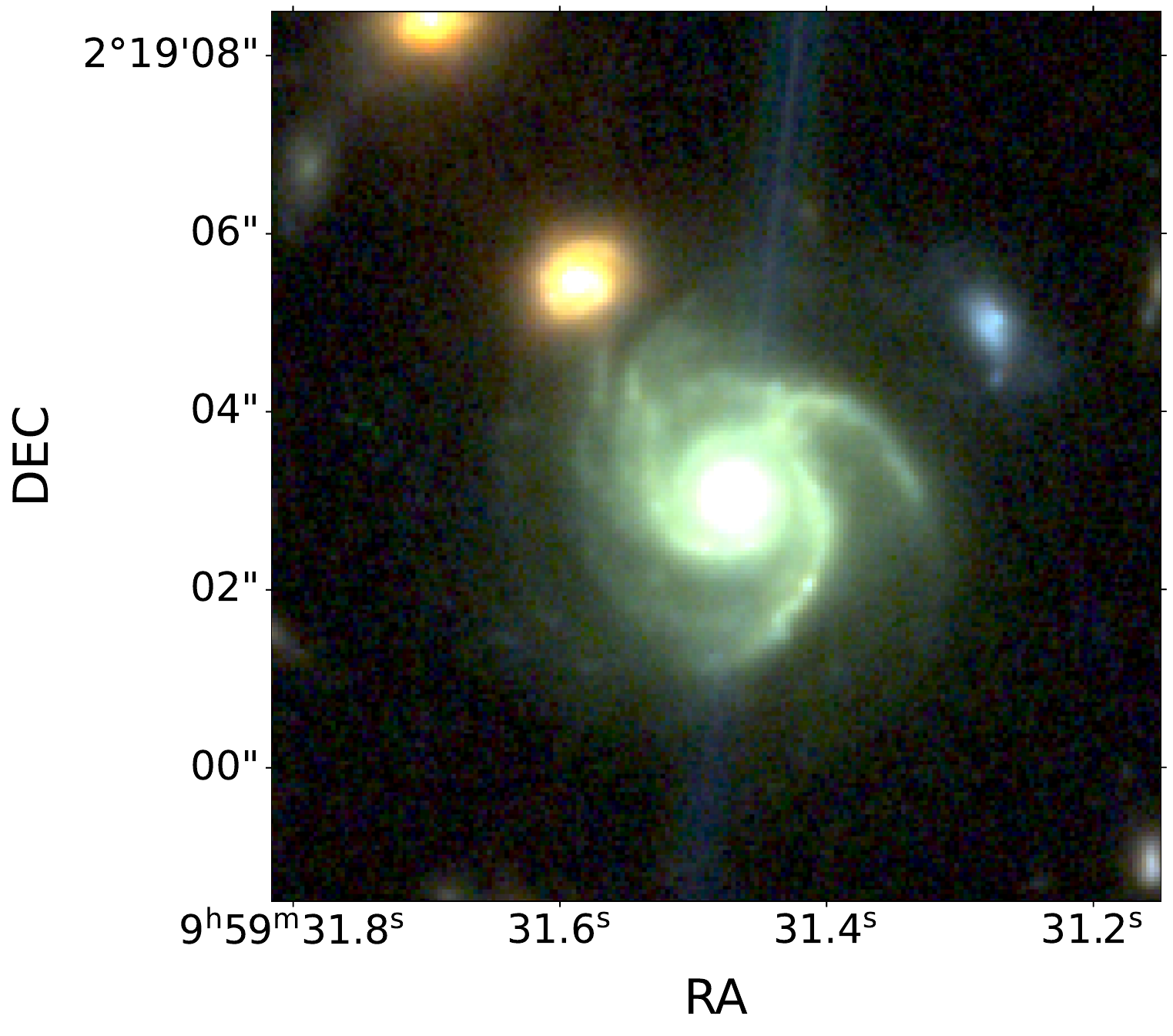}
     \end{subfigure}
     \hfill
     \begin{subfigure}[b]{0.24\textwidth}
         \centering
         \includegraphics[width=\textwidth]{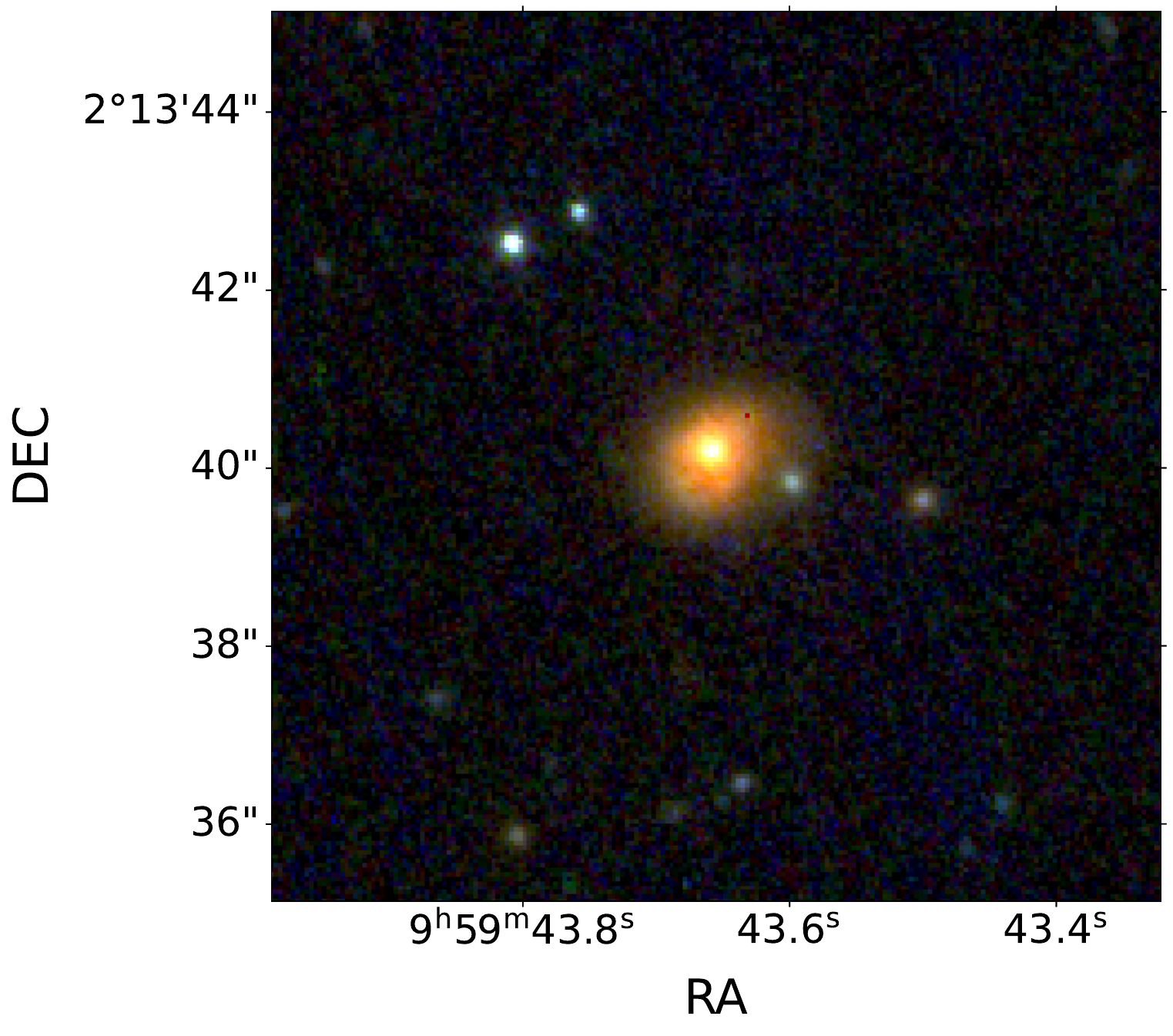}
     \end{subfigure}
     \hfill
     \begin{subfigure}[b]{0.24\textwidth}
         \centering
         \includegraphics[width=\textwidth]{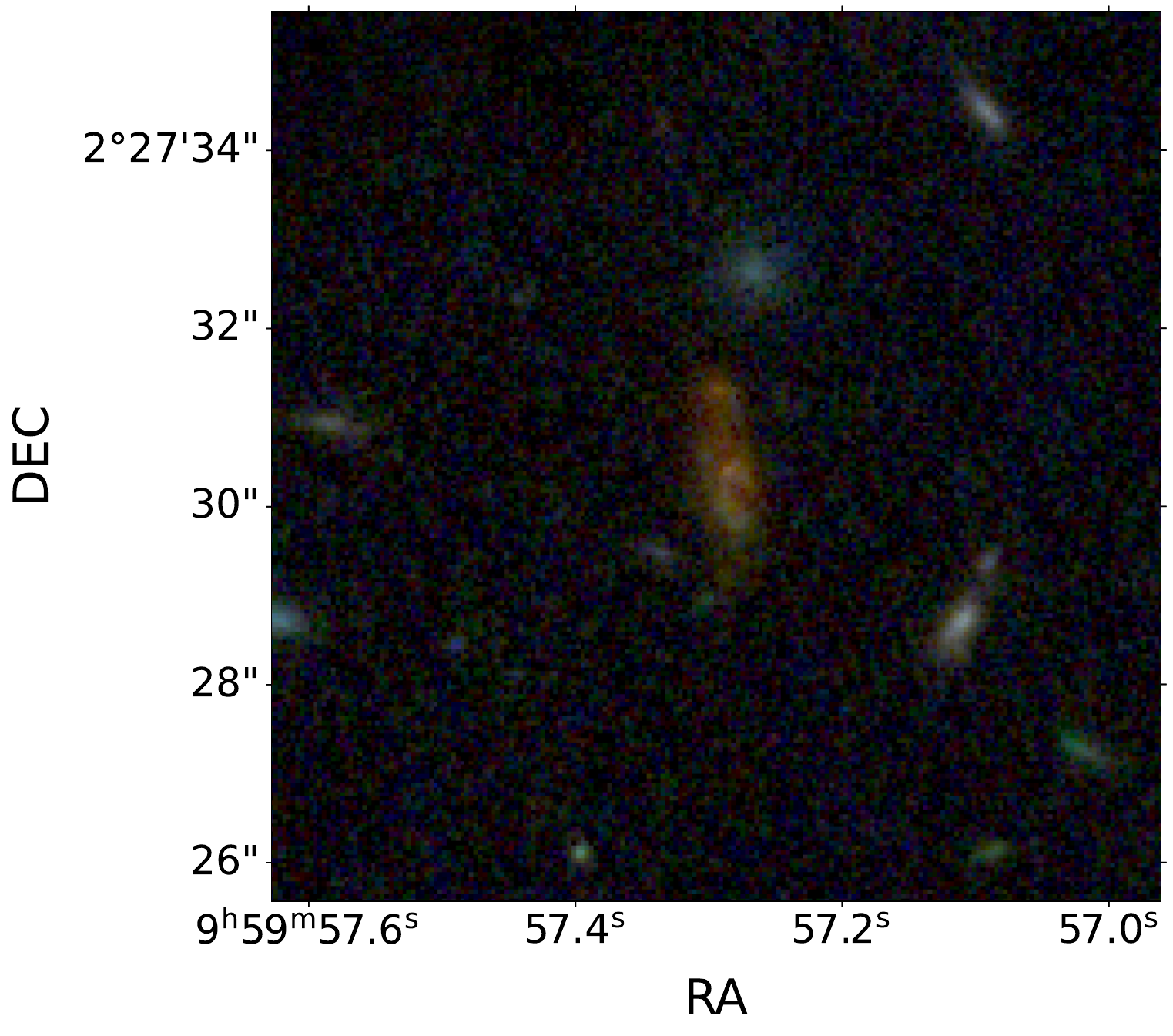}
     \end{subfigure}
     
     \caption{Composite color images of galaxies in the sample exhibiting different morphologies: merger, disk, compact, and irregular, respectively.}
     \label{fig: morpho}
\end{figure*}

Galaxies were classified into four categories: Disk, Compact, Irregular, and Merger, based on visual inspection of the sample's composite color NIRCam images. 13 team members independently repeated the inspection, and the modal classification was assigned to each galaxy. Out of the 52 galaxies in the study 38 were unanimously categorized. On the other 14 cases the agreement rate was always above 75\%.  Representative examples of each galaxy type are shown in Figure \ref{fig: morpho}.

\section{Molecular gas depending on galaxy morphology}\label{sect:morpho}

\begin{figure}[htbp]
     \centering
     \begin{subfigure}[b]{0.5\textwidth}
    \centering
    \includegraphics[width=1\linewidth]{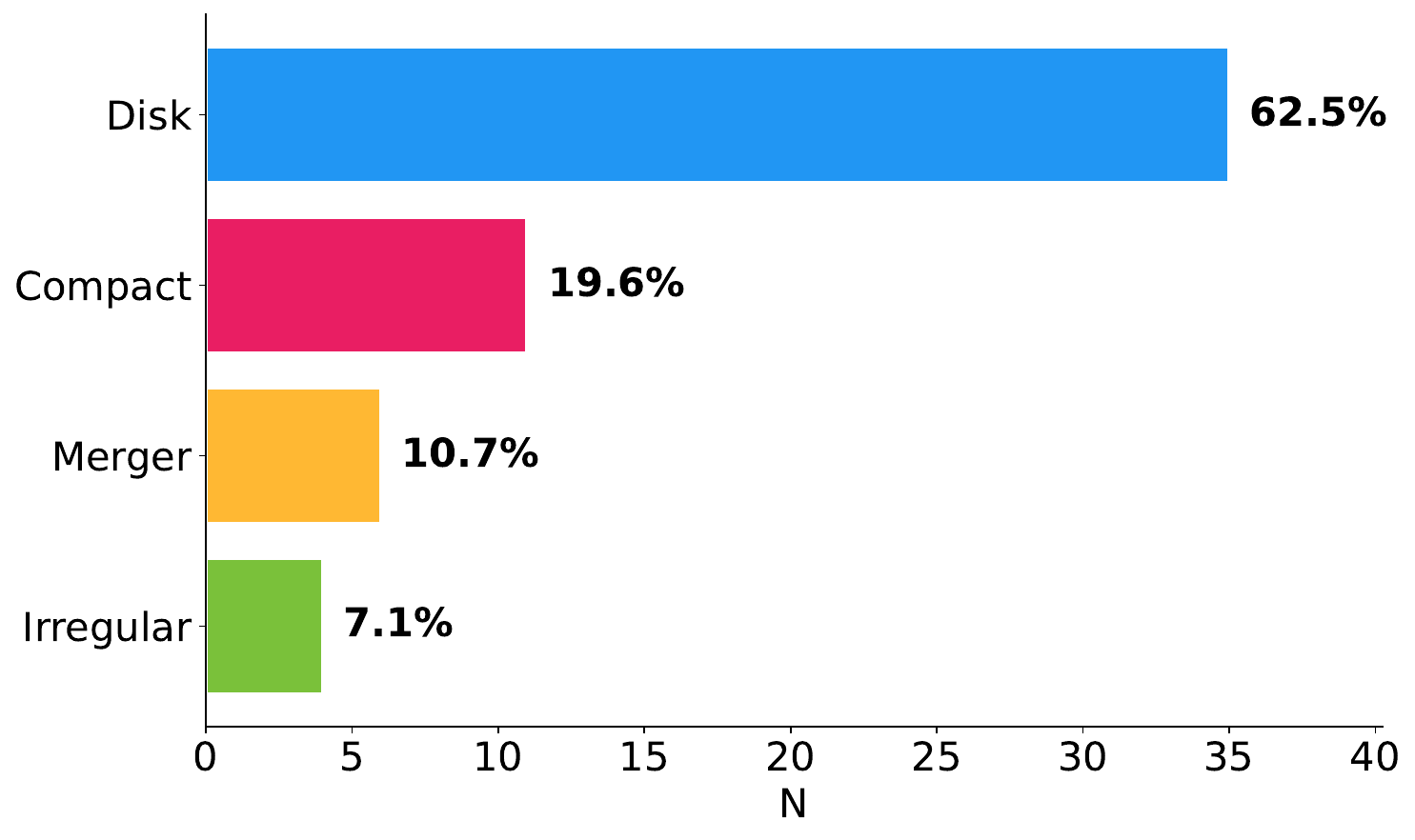}
    \caption{Classification of the different morphological types of our sample.}
    \label{fig: pie}
     \end{subfigure}
     \hfill 
     \begin{subfigure}[b]{0.45\textwidth}
        \centering
          \includegraphics[width=1\linewidth]{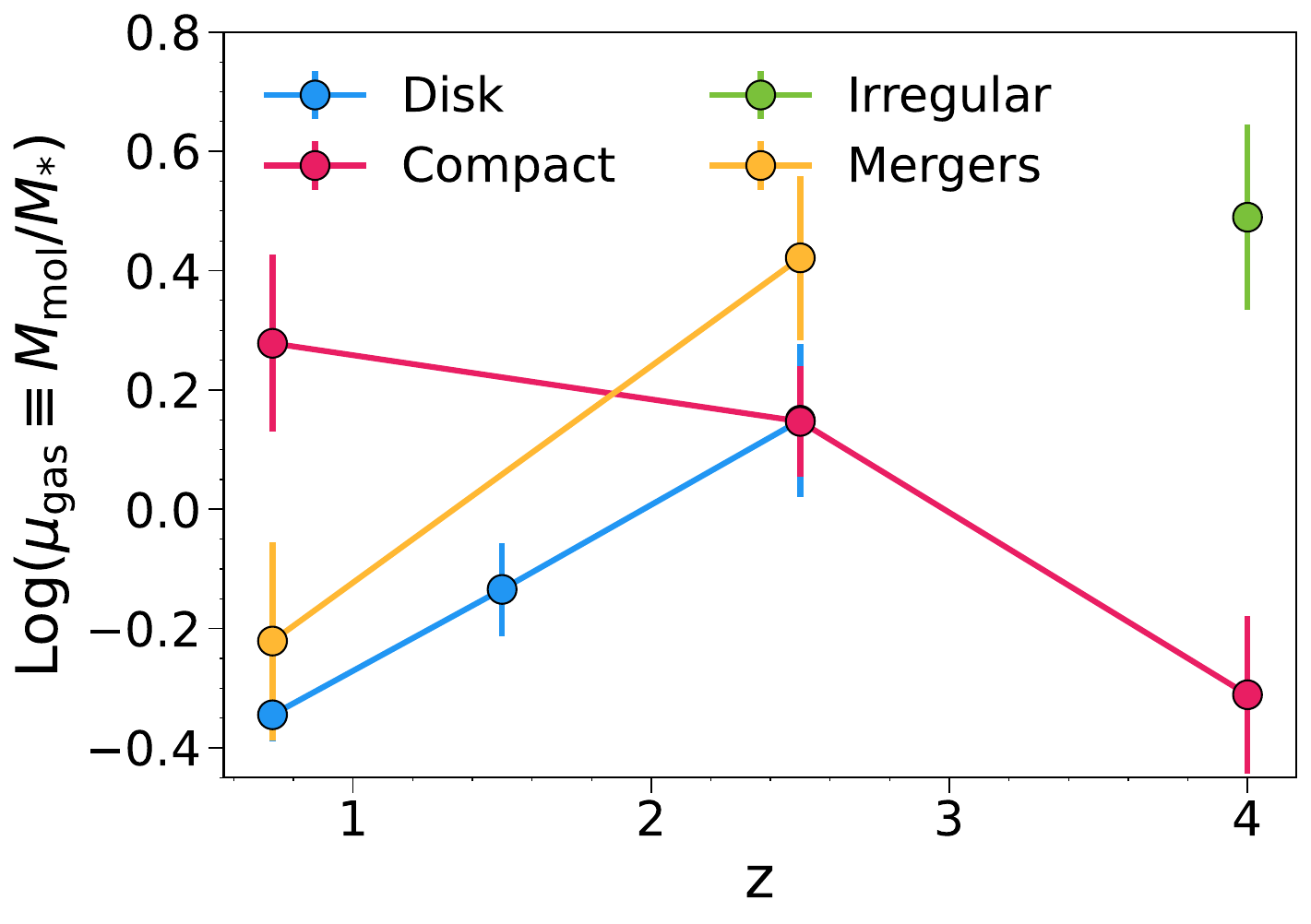}
          \caption{Evolution of average $\mu_{\text{gas}}$ as a function of $z$. Galaxies are categorized by their morphologies, with $1\sigma$ uncertainties indicated by vertical bars.}
    \label{fig: class_mu_morp}
     \end{subfigure}
\end{figure}

Galaxies were visually inspected and grouped into four morphological categories: Disk, Compact, Irregular, and Merger. Representative examples are presented in Figure~\ref{fig: morpho}. Figure~\ref{fig: pie} displays the distribution of each morphological type. The majority of the sample consists of disk galaxies, with a significant proportion of compact galaxies. This distribution supports the kinematical analysis presented in Section \ref{sect:kine}.

Figure~\ref{fig: class_mu_morp} illustrates the evolution of the average $\log(\mu_{\text{gas}})$ for each morphological type. Disk galaxies display an upward trend in $\log(\mu_{\text{gas}})$ with increasing redshift, while compact galaxies show a decrease. Mergers exhibit elevated $\log(\mu_{\text{gas}})$ values, peaking at $z \approx 2.5$. Irregular galaxies have the highest $\log(\mu_{\text{gas}})$ values in the highest-redshift bin, which is the only bin where they are identified. This trend may result from the increased difficulty in identifying morphological structures at higher redshifts, potentially leading to misclassification as irregular. The peak in mergers around the cosmic noon suggests that these galaxies have recently merged, leading to enhanced gas content within the sample.

\newpage
\section{Galaxy kinematics} \label{app:kine}
This section presents examples illustrating the kinematic classification of different galaxies within the sample.
\begin{figure*}[h]
    \centering
    \includegraphics[width=\linewidth]{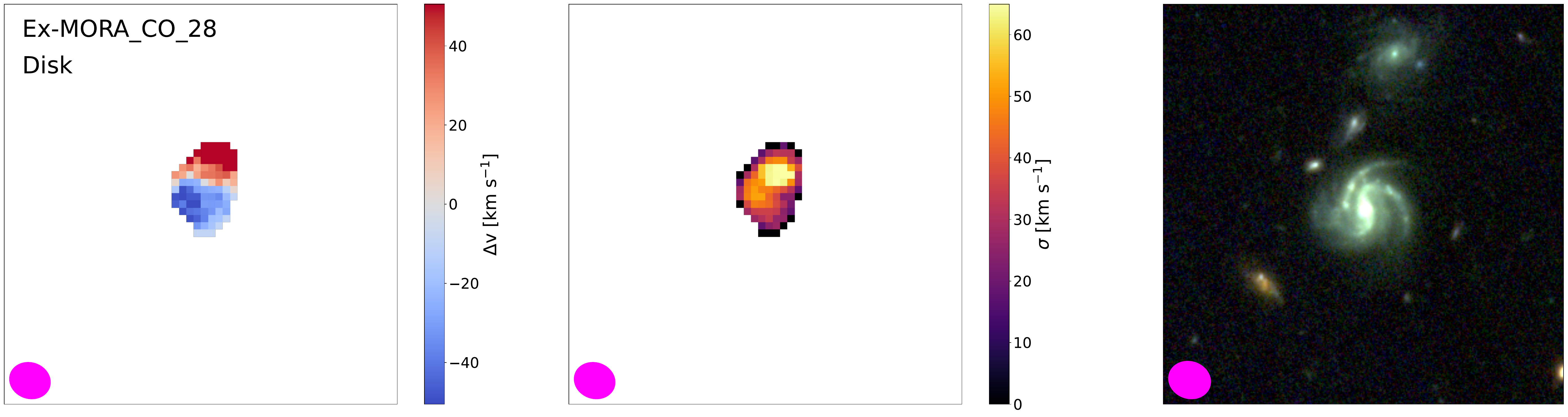} 
    \caption{Moment maps of a galaxy in the sample categorized as a disk. The left panel shows the first moment, the center panel displays the second moment, and the right panel presents a composite color image created from NIRCam data. The magenta oval in the bottom-left corner indicates the synthesized beam size}
    \label{fig:moment1}
\end{figure*}

\begin{figure*}[h]
    \centering
    \includegraphics[width=\linewidth]{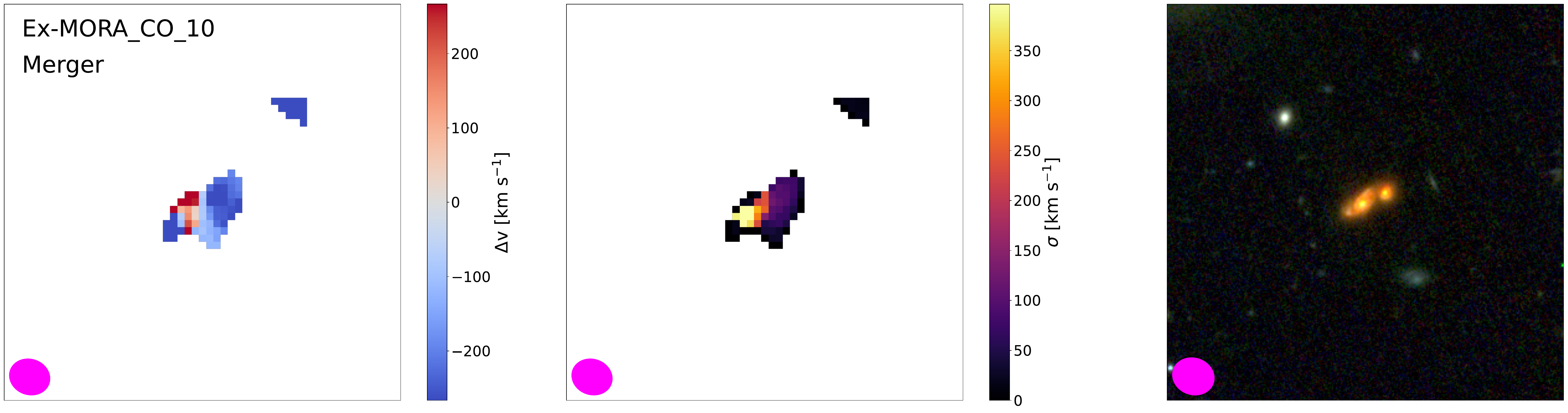}
    \caption{Moment maps of a galaxy in the sample categorized as a merger. The left panel shows the first moment, the center panel displays the second moment, and the right panel presents a composite color image created from NIRCam data. The magenta oval in the bottom-left corner indicates the synthesized beam size.}
    \label{fig:moment2}
\end{figure*}

\begin{figure*}[h]
    \centering
    \includegraphics[width=\linewidth]{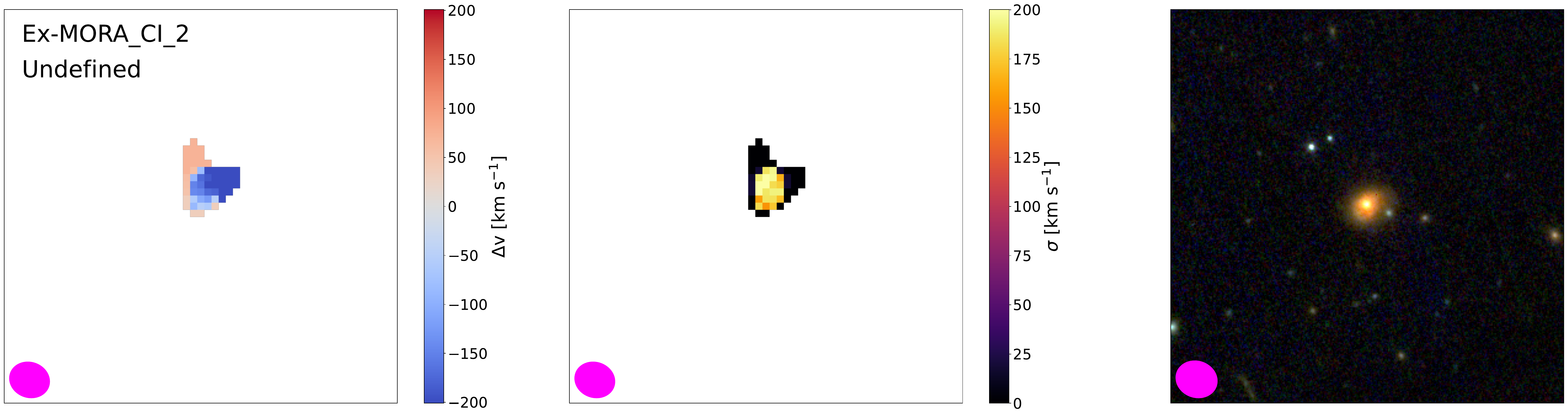}
    \caption{Moment maps of a galaxy in the sample categorized as undetermined. The left panel shows the first moment, the center panel displays the second moment, and the right panel presents a composite color image created from NIRCam data. The magenta oval in the bottom-left corner indicates the synthesized beam size.}
    \label{fig:moment3}
\end{figure*}

\section{Galaxy properties}\label{sec:table}
\begin{sidewaystable}
\centering
\caption{Galaxy properties} 
\centering
\begin{tabular}{l c c c c c c c c c }
\hline\hline
ID&RA&DEC&$z_{line}$&Line& Log(SFR) & Log(M$_*$)&$S_{line} \Delta \nu$&Log($L'_{Line}$)&Log($M_{mol}$)\\
&[deg]&[deg]&& [M$_{\odot}$yr$^{-1}$]  & [M$_{\odot}$]& [Jy km s$^{-1}$]& [K km s$^{-1}$ pc$^2$] & [M$_{\odot}$]\\
\hline
Ex-MORA\_CO\_1 & 150.12 & 2.534 & 3.3453(3) & CO(5-4) & 2.12$\pm$0.17 & 10.55$\pm$0.22 & 2.56$\pm$0.30 & 10.91$\pm$0.12 & 11.46$\pm$0.12 \\
Ex-MORA\_CO\_2 & 150.054 & 2.203 & 3.0908(1) & CO(5-4) & 2.28$\pm$0.29 & 11.32$\pm$0.06 & 1.48$\pm$0.20 & 10.62$\pm$0.12 & 11.01$\pm$0.12 \\
Ex-MORA\_CO\_3 & 150.105 & 2.313 & 2.29(1) & CO(4-3) & 2.67$\pm$0.13 & 10.82$\pm$0.12 & 0.93$\pm$0.20 & 10.28$\pm$0.13 & 10.77$\pm$0.13 \\
Ex-MORA\_CO\_4 & 150.086 & 2.298 & 1.2523(2) & CO(3-2) & 2.48$\pm$0.25 & 10.83$\pm$0.21 & 1.73$\pm$0.27 & 10.39$\pm$0.16 & 10.85$\pm$0.16 \\
Ex-MORA\_CO\_5 & 150.21 & 2.312 & 0.74729(7) & CO(2-1) & 2.27$\pm$0.24 & 10.66$\pm$0.14 & 1.05$\pm$0.13 & 9.97$\pm$0.08 & 10.44$\pm$0.08 \\
Ex-MORA\_CO\_6 & 150.147 & 2.337 & 0.7262(2) & CO(2-1) & 1.93$\pm$0.37 & 10.94$\pm$0.15 & 1.56$\pm$0.16 & 10.12$\pm$0.08 & 10.58$\pm$0.08 \\
Ex-MORA\_CO\_7 & 149.953 & 2.516 & 0.7275(1) & CO(2-1) & 1.78$\pm$0.60 & 10.99$\pm$0.16 & 2.16$\pm$0.21 & 10.26$\pm$0.08 & 10.72$\pm$0.08 \\
Ex-MORA\_\ci\_1 & 149.922 & 2.29 & 2.3703(4) & \ci (1-0) & 2.59$\pm$0.30 & 11.12$\pm$0.25 & 3.80$\pm$0.50 & 10.74$\pm$0.06 & 11.90$\pm$0.10 \\
Ex-MORA\_CO\_8 & 150.237 & 2.338 & 2.4946(3) & CO(4-3) & 2.41$\pm$0.32 & 10.89$\pm$0.29 & 3.12$\pm$0.22 & 10.87$\pm$0.10 & 11.35$\pm$0.10 \\
Ex-MORA\_CO\_9 & 149.881 & 2.318 & 0.73212(5) & CO(2-1) & 0.81$\pm$2.02 & 11.25$\pm$0.13 & 0.57$\pm$0.13 & 9.69$\pm$0.11 & 10.15$\pm$0.11 \\
Ex-MORA\_\ci\_2 & 149.932 & 2.228 & 2.6800(7) & \ci (1-0) & 2.57$\pm$0.38 & 11.41$\pm$0.12 & 0.62$\pm$0.13 & 10.04$\pm$0.09 & 11.10$\pm$0.10 \\
Ex-MORA\_CO\_10 & 150.02 & 2.513 & 2.4766(6) & CO(4-3) & 2.77$\pm$0.27 & 11.22$\pm$0.27 & 3.63$\pm$0.29 & 10.93$\pm$0.10 & 11.34$\pm$0.10 \\
Ex-MORA\_CO\_11 & 150.201 & 2.461 & 1.58221(8) & CO(3-2) & 2.49$\pm$0.15 & 10.67$\pm$0.16 & 0.88$\pm$0.15 & 10.34$\pm$0.11 & 10.80$\pm$0.11 \\
Ex-MORA\_\ci\_3 & 150.143 & 2.356 & 2.4056(2) & \ci (1-0) & 1.62$\pm$0.52 & 10.51$\pm$0.23 & 0.57$\pm$0.31 & 9.93$\pm$0.24 & 11.29$\pm$0.25 \\
Ex-MORA\_CO\_12 & 149.88 & 2.243 & 0.61(1) & CO(2-1) & 2.46$\pm$0.32 & 10.78$\pm$0.21 & 0.63$\pm$0.25 & 9.57$\pm$0.18 & 10.03$\pm$0.18 \\
Ex-MORA\_\ci\_4 & 150.137 & 2.514 & 2.3969(4) & \ci (1-0) & 2.03$\pm$0.84 & 11.53$\pm$0.09 & 2.34$\pm$0.47 & 10.54$\pm$0.09 & 11.56$\pm$0.09 \\
Ex-MORA\_CO\_13 & 150.056 & 2.374 & 0.59373(5) & CO(2-1) & 1.26$\pm$0.09 & 10.33$\pm$0.22 & 2.39$\pm$0.62 & 10.11$\pm$0.19 & 10.57$\pm$0.19 \\
Ex-MORA\_\ci\_5 & 149.919 & 2.345 & 2.4377(7) & \ci (1-0) & 1.59$\pm$0.36 & 10.02$\pm$0.42 & 2.31$\pm$0.47 & 10.55$\pm$0.09 & 12.08$\pm$0.17 \\
Ex-MORA\_CO\_14 & 150.083 & 2.536 & 1.4204(4) & CO(3-2) & 2.64$\pm$0.14 & 11.02$\pm$0.12 & 1.27$\pm$0.22 & 10.41$\pm$0.11 & 10.87$\pm$0.11 \\
Ex-MORA\_\ci\_6 & 150.057 & 2.293 & 2.43(2) & \ci (1-0) & 1.40$\pm$0.59 & 11.35$\pm$0.07 & 0.53$\pm$0.14 & 9.91$\pm$0.12 & 10.98$\pm$0.12 \\
Ex-MORA\_\ci\_7 & 150.069 & 2.444 & 2.470(1) & \ci (1-0) & 2.46$\pm$0.40 & 11.28$\pm$0.13 & 0.85$\pm$0.22 & 10.12$\pm$0.11 & 11.22$\pm$0.12 \\
Ex-MORA\_CO\_15 & 150.073 & 2.233 & 2.32(2) & CO(4-3) & 0.55$\pm$0.75 & 10.91$\pm$0.07 & 0.66$\pm$0.15 & 10.45$\pm$0.17 & 10.92$\pm$0.17 \\
Ex-MORA\_CO\_16 & 150.042 & 2.526 & 1.2470(2) & CO(3-2) & 1.38$\pm$0.73 & 11.16$\pm$0.09 & 2.27$\pm$0.44 & 10.55$\pm$0.11 & 11.01$\pm$0.11 \\
Ex-MORA\_CO\_17 & 150.079 & 2.303 & 0.5023(1) & CO(2-1) & 1.46$\pm$0.26 & 10.43$\pm$0.15 & 0.97$\pm$0.18 & 9.58$\pm$0.10 & 10.05$\pm$0.10 \\
Ex-MORA\_CO\_18 & 149.961 & 2.218 & 0.7332(3) & CO(2-1) & 2.04$\pm$0.50 & 10.85$\pm$0.17 & 0.93$\pm$0.14 & 9.89$\pm$0.17 & 10.35$\pm$0.17 \\
Ex-MORA\_CO\_19 & 150.133 & 2.262 & 0.7507(3) & CO(2-1) & 1.64$\pm$0.36 & 10.22$\pm$0.20 & 0.99$\pm$0.14 & 9.95$\pm$0.09 & 10.42$\pm$0.09 \\
Ex-MORA\_CO\_20 & 150.165 & 2.289 & 0.7265(2) & CO(2-1) & 1.79$\pm$0.18 & 10.35$\pm$0.28 & 1.03$\pm$0.13 & 9.94$\pm$0.08 & 10.40$\pm$0.08 \\
Ex-MORA\_CO\_21 & 150.178 & 2.292 & 0.75(3) & CO(2-1) & 2.03$\pm$0.32 & 10.57$\pm$0.21 & 0.80$\pm$0.21 & 9.86$\pm$0.13 & 10.32$\pm$0.13 \\
Ex-MORA\_CO\_22 & 149.883 & 2.331 & 0.7335(2) & CO(2-1) & 1.47$\pm$0.34 & 10.47$\pm$0.20 & 0.33$\pm$0.10 & 9.46$\pm$0.14 & 9.92$\pm$0.14 \\
\hline
\hline
\end{tabular}
\tablefoot{Derived values include Star Formation Rate (SFR) and stellar mass ($M_*$) from \texttt{CIGALE} SED modeling, alongside molecular gas properties including integrated line flux ($S_{line}\Delta\nu$), line luminosity ($L'_{Line}$), and total molecular gas mass ($M_{mol}$). The number in parenthesis for $z_{line}$ corresponds to the error in the last decimal.}
\label{table:properties2}
\end{sidewaystable}
\begin{sidewaystable}
\centering
\caption{Continuation Table \ref{table:properties2}} 
\centering
\begin{tabular}{l c c c c c c c c c }
\hline\hline
ID&RA&DEC&$z_{line}$&Line& Log(SFR) & Log(M$_*$)&$S_{line} \Delta \nu$&Log($L'_{Line}$)&Log($M_{mol}$)\\
&[deg]&[deg]&& [M$_{\odot}$yr$^{-1}$]  & [M$_{\odot}$]& [Jy km s$^{-1}$]& [K km s$^{-1}$ pc$^2$] & [M$_{\odot}$]\\
\hline
Ex-MORA\_CO\_23 & 149.942 & 2.396 & 0.77(1) & CO(2-1) & 2.13$\pm$0.37 & 11.13$\pm$0.15 & 0.56$\pm$0.13 & 9.73$\pm$0.12 & 10.19$\pm$0.12 \\
Ex-MORA\_CO\_24 & 149.963 & 2.442 & 0.73(3) & CO(2-1) & 1.51$\pm$0.73 & 10.91$\pm$0.14 & 0.42$\pm$0.11 & 9.56$\pm$0.13 & 10.02$\pm$0.13 \\
Ex-MORA\_CO\_25 & 149.85 & 2.452 & 0.7147(1) & CO(2-1) & 2.04$\pm$0.32 & 10.53$\pm$0.15 & 1.00$\pm$0.20 & 9.90$\pm$0.18 & 10.36$\pm$0.18 \\
Ex-MORA\_CO\_26 & 149.994 & 2.449 & 0.7318(2) & CO(2-1) & 0.90$\pm$0.55 & 11.09$\pm$0.11 & 1.55$\pm$0.17 & 10.13$\pm$0.08 & 10.59$\pm$0.08 \\
Ex-MORA\_CO\_27 & 149.854 & 2.468 & 0.7308(1) & CO(2-1) & 1.60$\pm$0.54 & 10.69$\pm$0.16 & 1.84$\pm$0.22 & 10.20$\pm$0.08 & 10.66$\pm$0.08 \\
Ex-MORA\_CO\_28 & 149.99 & 2.491 & 0.7334(1) & CO(2-1) & 0.89$\pm$0.26 & 9.31$\pm$0.20 & 0.47$\pm$0.13 & 9.61$\pm$0.14 & 10.07$\pm$0.14 \\
Ex-MORA\_CO\_29 & 149.915 & 2.506 & 0.7244(1) & CO(2-1) & 1.62$\pm$0.35 & 10.92$\pm$0.13 & 0.87$\pm$0.13 & 9.86$\pm$0.09 & 10.32$\pm$0.09 \\
Ex-MORA\_CO\_30 & 149.88 & 2.513 & 0.73038(5) & CO(2-1) & 1.50$\pm$0.33 & 10.71$\pm$0.14 & 0.65$\pm$0.13 & 9.74$\pm$0.11 & 10.20$\pm$0.11 \\
Ex-MORA\_CO\_31 & 149.874 & 2.519 & 0.73485(8) & CO(2-1) & 1.13$\pm$0.38 & 10.74$\pm$0.11 & 0.98$\pm$0.16 & 9.93$\pm$0.10 & 10.39$\pm$0.10 \\
Ex-MORA\_CO\_32 & 149.987 & 2.549 & 0.73196(6) & CO(2-1) & 1.90$\pm$0.50 & 10.84$\pm$0.18 & 0.59$\pm$0.11 & 9.71$\pm$0.10 & 10.17$\pm$0.10 \\
Ex-MORA\_CO\_33 & 149.986 & 2.554 & 0.7269(3) & CO(2-1) & 2.01$\pm$0.40 & 10.78$\pm$0.17 & 1.43$\pm$0.14 & 10.08$\pm$0.08 & 10.54$\pm$0.08 \\
Ex-MORA\_CO\_34 & 150.055 & 2.431 & 0.7291(3) & CO(2-1) & 1.95$\pm$0.36 & 10.79$\pm$0.22 & 2.71$\pm$0.37 & 10.36$\pm$0.09 & 10.82$\pm$0.09 \\
Ex-MORA\_CO\_35 & 150.096 & 2.16 & 0.72571(9) & CO(2-1) & 1.22$\pm$0.35 & 10.74$\pm$0.12 & 1.49$\pm$0.16 & 10.10$\pm$0.08 & 10.56$\pm$0.08 \\
Ex-MORA\_CO\_36 & 150.011 & 2.452 & 0.737(9) & CO(2-1) & 1.65$\pm$0.33 & 10.20$\pm$0.18 & 0.66$\pm$0.16 & 9.76$\pm$0.12 & 10.22$\pm$0.12 \\
Ex-MORA\_CO\_37 & 149.919 & 2.501 & 0.73(2) & CO(2-1) & 1.20$\pm$0.20 & 9.75$\pm$0.39 & 0.79$\pm$0.14 & 9.83$\pm$0.10 & 10.30$\pm$0.10 \\
Ex-MORA\_CO\_38 & 149.882 & 2.318 & 1.6167(2) & CO(3-2) & 1.83$\pm$0.34 & 11.18$\pm$0.12 & 0.72$\pm$0.14 & 10.27$\pm$0.11 & 10.73$\pm$0.11 \\
Ex-MORA\_CO\_39 & 150.15 & 2.364 & 2.4639(3) & CO(4-3) & 2.20$\pm$0.34 & 10.80$\pm$0.25 & 1.88$\pm$0.26 & 10.64$\pm$0.11 & 11.14$\pm$0.11 \\
Ex-MORA\_CO\_40 & 150.139 & 2.434 & 2.513(4) & CO(4-3) & 2.58$\pm$0.26 & 11.08$\pm$0.22 & 1.31$\pm$0.26 & 10.50$\pm$0.12 & 10.93$\pm$0.12 \\
Ex-MORA\_CO\_41 & 150.069 & 2.444 & 2.472(1) & CO(4-3) & 2.59$\pm$0.27 & 11.06$\pm$0.26 & 0.90$\pm$0.19 & 10.62$\pm$0.17 & 11.06$\pm$0.17 \\
Ex-MORA\_CO\_42 & 150.24 & 2.336 & 2.5134(5) & CO(4-3) & 2.57$\pm$0.29 & 11.10$\pm$0.14 & 0.92$\pm$0.14 & 10.35$\pm$0.11 & 10.78$\pm$0.11 \\
Ex-MORA\_CO\_43 & 150.034 & 2.437 & 4.6197(6) & CO(7-6) & 2.82$\pm$0.38 & 11.16$\pm$0.28 & 1.47$\pm$0.35 & 11.09$\pm$0.23 & 11.51$\pm$0.23 \\
Ex-MORA\_CO\_44 & 150.065 & 2.264 & 4.6013(3) & CO(7-6) & 2.68$\pm$0.36 & 11.06$\pm$0.22 & 1.10$\pm$0.20 & 10.97$\pm$0.22 & 11.41$\pm$0.22 \\
Ex-MORA\_\ci\_8 & 150.034 & 2.437 & 4.64(3) & CI2-1 & 2.82$\pm$0.38 & 11.16$\pm$0.28 & 0.63$\pm$0.16 & 10.36$\pm$0.11 & 11.50$\pm$0.14 \\
Ex-MORA\_CO\_45 & 150.228 & 2.47 & 0.6044(2) & CO(2-1) & 1.73$\pm$0.42 & 10.64$\pm$0.19 & 2.27$\pm$0.28 & 10.11$\pm$0.16 & 10.57$\pm$0.16 \\
Ex-MORA\_CO\_46 & 150.131 & 2.361 & 1.40(2) & CO(3-2) & 2.23$\pm$0.27 & 10.78$\pm$0.17 & 0.66$\pm$0.15 & 10.11$\pm$0.13 & 10.57$\pm$0.13 \\
\hline
\hline
\end{tabular}
\tablefoot{Derived values include Star Formation Rate (SFR) and stellar mass ($M_*$) from \texttt{CIGALE} SED modeling, alongside molecular gas properties including integrated line flux ($S_{line}\Delta\nu$), line luminosity ($L'_{Line}$), and total molecular gas mass ($M_{mol}$). The number in parenthesis for $z_{line}$ corresponds to the error in the last decimal.}
\label{table:properties3}
\end{sidewaystable}
\end{appendix}
\
\end{document}